\documentclass[conference]{IEEEtran}
\IEEEoverridecommandlockouts
\usepackage{cite}
\usepackage{amsmath,amssymb,amsfonts}
\usepackage{algorithmic}
\usepackage{graphicx}
\usepackage{textcomp}
\usepackage{xcolor}
\usepackage{subfig}
\usepackage{subcaption}

\DeclareMathOperator*{\argmin}{arg\,min}

\def\BibTeX{{\rm B\kern-.05em{\sc i\kern-.025em b}\kern-.08em
    T\kern-.1667em\lower.7ex\hbox{E}\kern-.125emX}}
\begin{document}

\newpage
\thispagestyle{empty}

\begin{center}
    \textbf{Copyright © 2023 IEEE. Personal use of this material is permitted.  Permission from IEEE must be obtained for all other uses, in any current or future media, including reprinting/republishing this material for advertising or promotional purposes, creating new collective works, for resale or redistribution to servers or lists, or reuse of any copyrighted component of this work in other works.}
\end{center}

\title{Performance Tuning for GPU-Embedded Systems: Machine-Learning-based and Analytical Model-driven Tuning Methodologies.}

 \author{\IEEEauthorblockN{1\textsuperscript{st} Adri\'an P. Di\'eguez}
 \IEEEauthorblockA{\textit{Lawrence Berkeley National Laboratory} \\
 Berkeley, CA, USA \\
 aperezdieguez@lbl.gov}
 \and
 \IEEEauthorblockN{2\textsuperscript{nd} Margarita Amor L\'opez}
 \IEEEauthorblockA{\textit{University of A Coru\~na} \\
 A Coru\~na, Spain \\
  margamor@udc.es}
% \and
 }
 
\maketitle

\begin{abstract}

GPU-embedded systems have gained popularity across various domains due to their efficient power consumption. However, in order to meet the demands of real-time or time-consuming applications running on these systems, it is crucial for them to be tuned to exhibit high performance. This paper addresses the issue by developing and comparing two tuning methodologies on GPU-embedded systems, and also provides performance insights for developers and researchers seeking to optimize applications running on these architectures. We focus on parallel prefix operations, such as FFT, scan primitives, and tridiagonal system solvers, which are performance-critical components in many applications. The study introduces an analytical model-driven tuning methodology and a Machine Learning (ML)-based tuning methodology. We evaluate the performance of the two tuning methodologies for different parallel prefix implementations of the BPLG library in an NVIDIA Jetson system, and compare their performance to the ones achieved through an exhaustive search. The findings shed light on the best strategies for handling the open challenge of performance portability for major computational patterns, providing practical guidance for offline and online tuning. We also address the existing gap in performance studies for parallel computational patterns in GPU-embedded systems by comparing the BPLG performance against other state-of-the-art libraries, including CUSPARSE, CUB, and CUFFT.

\end{abstract}

%\begin{IEEEkeywords}
% TBC
%\end{IEEEkeywords}

\section{Introduction}

In the realm of HPC, performance portability has emerged as a paramount concern as applications are increasingly deployed across diverse computing platforms. The ability to achieve optimal performance regardless of the underlying architecture is crucial for attaining efficient utilization of computational resources and maximizing the scalability of applications. This need for performance portability has exacerbated even more pronounced with the rise of heterogeneous systems. GPU-embedded systems combine high-performance computing capabilities with embedded architectures, which have found extensive applications in domains such as image processing, motion planning, signal processing and the Internet of Things (IoT). Unfortunately, not many performance studies have been published about GPU-embedded platforms, earlier studies~\cite{7939053} show that choosing the optimal performance parameters for different computation patterns running on an embedded GPU can be challenging. 

Auto-tuning may be applied to tackle this issue. This strategy  efficiently explore a subset of the search space of all possible configurations through empirical measurements and/or predictive models to identify the best runtime configuration. An \textit{exhaustive search}, which evaluates all configurations, guarantees finding the optimal configuration, but can be infeasible for modern applications. Meanwhile, a \textit{predictive search} can be divided into the traditional \textit{analytical model-driven search} and a \textit{Machine-Learning-based search}. In the former, an expert builds the model with performance heuristics, but results are strongly correlated to the quality of the performance model. In the later, the relation between performance and tuning parameters can be cast as a black-box model, but sampling the whole space for getting training data can be expensive. Finding the best strategy for performance portability is still an open challenge for many computational patterns. For instance, \textit{Parallel prefix operations} \cite{Ladner:1980:PPC:322217.322232}, such as FFT, the scan primitive and tridiagonal system solvers, are an example of performance-critical components used in many applications. Showing a methodology for tuning these kernels to GPU-embedded platforms would help developers and researchers to optimize their applications for these platforms.

This work has two main contributions: it compares the efficiency of two tuning methodologies on embedded systems for performance portability, an open challenge in computer science, and evaluates the performance of different state-of-the-art libraries applied to different parallel prefix operations on an NVIDIA Jetson system, which lacks performance benchmarks for the explored kernels. For achieving the first objective, this work uses the existing BPLG library \cite{BPLGJacobo} where performance parameters for the FFT, the scan primitive and tridiagonal-system solver implementations are exposed as template arguments to be defined by users for each problem size and operation. In order to tune these values, we developed two tuning methodologies: an analytical model-driven tuning methodology based on \cite{jos2018} but extended to GPU-embedded systems; and a ML-based tuning methodology, based on Bayesian optimization (BO). To target our second goal, we developed a benchmark to evaluate the performance achieved with the configurations proposed by each methodology, comparing them with state-of-the-art libraries such as CUSPARSE, CUB, and CUFFT.

\subsection{Related Work}

 Regarding performance studies about GPU-embedded systems, previous works, such as those in \cite{7336442, franciscomaster, 7307646}, have provided valuable performance insights into the Jetson Tegra TK1. In \cite{pepe}, a performance characterization of both the Jetson TK1 and TX1 is conducted using roofline models, focusing on a matrix-multiplication application. However, the simplicity of the matrix multiplication fails to capture the complexities of real-world workloads. In \cite{algomas, 10.1145/3570604, 9882480}, machine learning applications on the Jetson TX1, Nano, and AGX Xavier are respectively explored, offering significant performance insights but with a narrow focus on specific machine learning tasks. While some works have emphasized energy efficiency, comprehensive performance benchmarking has been lacking, except for \cite{10.1145/3434770.3459729}, which also includes a performance variability analysis for the Jetson AGX Xavier. Notably, few studies have focused on performance, and the existing ones mostly rely on benchmarking approaches rather than developing performance models that can be applied to other embedded platforms.
 
In the field of desktop and server GPUs, there are examples of empirical tuning projects for the FFT such as \cite{article1234,YURIFFT,NUKADA, NUKADA2, Park13,6755214, 9555937}. GPU tridiagonal system solvers (TS)  have been proposed in  \cite{Davidson:2011:RPF} and \cite{DBLP:conf/icpp/KimWCH11}. Nevertheless, these TS approaches do not exhibit significant optimization for modern architectures. 
 More recent GPU TS can be found in \cite{Chang:2012:SNS:2388996.2389033, rtrTridiagonales,TRIDIAG4,Laszlo:2016:MAB:2956571.2830568,Yang:2017:PSM:3092142.3092210,9919397}, while \cite{Dotsenko:2008:FSA:1375527.1375559,Yan:2013:SFS:2517327.2442539,jos2018} introduce optimized GPU implementations for the scan primitive. Some tuning libraries, such as \textit{NVIDIA's} \textit{CUFFT} \cite{CUFFT} for FFT, \textit{CUSPARSE} \cite{cusparse} for tridiagonal system solvers, and  the  \textit{CUB} library \cite{CUBlib} for the scan primitive, provide efficient implementations and are considered the state-of-the-art for these operations. In the realm of server GPU architectures, performance modeling has been explored  in \cite{jos2018}, while we extend their outlined principles to the context of GPU-embedded systems in this work, as detailed in the subsequent sections.

With respect to generic autotuning, where the goal is to tune any objective function rather than focusing on specific algorithms or platforms, we can find autotuners based on empirical searches such as OpenTuner \cite{KERNEL13} that use empirical approaches, being very slow. Some other generic auto-tuners, such as Kernel Tuner \cite{KERNEL}, implement various search optimization algorithms, but most of them have been demonstrated to not be more efficient than a random search \cite{KERNEL16}. Other research efforts have developed machine learning models  \cite{OPENCL30,OPENCL31,OpenCL, cox},  but using random sampling indiscriminately to build the surrogate model. Instead, Bayesian optimization identifies the most informative candidates to train the surrogate model and achieves the same accuracy with fewer samples \cite{active} \cite{BO21}. Two generic auto-tuners, GPTune \cite{gptune} and DeepHyper \cite{Dorier_2022} use Bayesian optimization and are recognized as cutting-edge autotuners for HPC applications.

\section{The GPU-Embedded Architecture}
\label{jetsondes}

This work uses a NVIDIA Jetson TX1 as underlying platform for testing the proposed tuning methodologies on GPU-embedded platforms. While the NVIDIA Jetson TX1 may not be the latest architecture from NVIDIA, it can still serve as a relevant and valuable target platform for our studies: (i) it gained significant popularity when it was first released, and it still continues to have a strong user base in various applications and industries; (ii) The Jetson TX1 offers a cost-effective solution for GPU-embedded development, its affordability makes it accessible to a wide range of users; and (iii) while it may not have the latest capabilities of newer architectures, it can still provide a useful baseline for performance comparison and by demonstrating optimizations on this platform, one can highlight the potential gains on more recent architectures. Therefore, the methodologies developed for performance tuning on the Jetson TX1 can be used on other GPU-embedded platforms, allowing this work to benefit a wider audience interested in performance tuning and optimization in the context of GPU-embedded computing.

\begin{figure}[t!]
\centering
\includegraphics[scale=0.55]{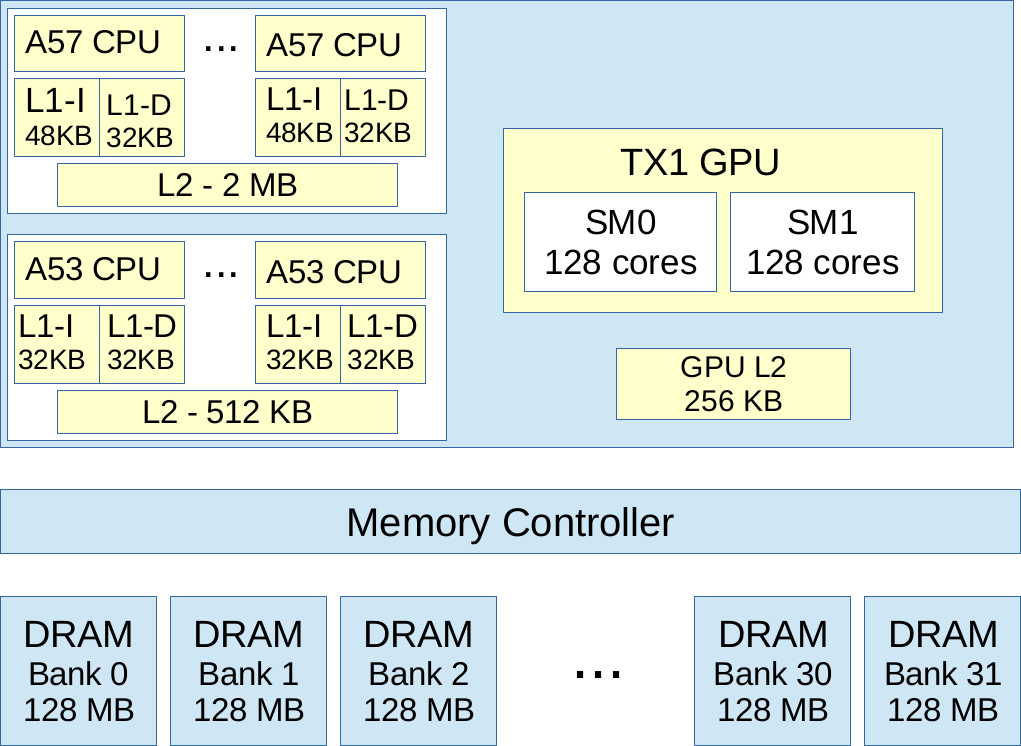}
\caption{NVIDIA's Jetson TX1 architecture design}
\label{tx1-arch}
\end{figure}

The NVIDIA Jetson TX1 module has a GM20B Maxwell GPU with 256 cores in addition to four ARM Cortex-A57 and four Cortex-A53 cores. They share 4 GB of LPDDR4 memory and 16 GB of eMMC flash storage. It has 1 Gigabit Ethernet interface and six CSI cameras interfaces. It also supports wireless with 802.11ac Wi-Fi and Bluetooth. Specifically, the SoC (system-on-chip) contains a quad-core 1.91 GHz 64-bit ARM Cortex-A57 with an integrated GM20B NVIDIA GPU \cite{tegra}. This GPU has two Streaming Multiprocessor (SM), each with 128 CUDA cores, providing up to 512 GFLOPS in single precision. It also contains a 256-KB L2 cache; whereas the A57 CPU has a 2-MB L2 cache. Although this board contains a second-generation Maxwell GPU, it should be observed that it only has 2 SM and 256-KB L2 cache, in comparison to the 16 SMs and 2048-KB L2 cache, as well as the high bandwidth, of a second-generation Maxwell GeForce card. This architecture also provides 64 KB of shared memory per SM, although a maximum of 48 KB per threadblock. The maximum number of warps per SM is 64, and the register file contains 65536 simple-precision registers; but only 32 warps and 32768 registers, respectively, can be used at most per threadblock. The total number of threadblocks executed in each SM is 32. Finally, the TX1 module also contains a quad-core ARM Cortex A53 CPU, which is not directly accessible to software. The module shares 4 GB of 1600-MHz DRAM memory partitioned into 32 banks. The SoC architecture is showed in Figure \ref{tx1-arch}.

\section{Parallel Prefix Operations on a GPU}
\label{parallelprefix}

Parallel prefix algorithms, which are well-suited for GPU architecture, exhibit regular characteristics. These algorithms have static communication patterns that can be expressed as linear functions with the element index as a variable. Additionally, the resulting elements are computed based on combinations of other elements. A parallel prefix algorithm \cite{Ladner:1980:PPC:322217.322232}  solves a problem of size $N=r^n$  in $K$ steps, where $r$ is a power of two called \textit{radix}.  The parallel prefix algorithm  is depicted by a directed acyclic oriented graph called prefix circuit. The computations are performed by the \textit{Node operator}, responsible for executing the required operation on the given elements. This operator is represented by small circles in  prefix circuits shown in Figure \ref{fig:parallelprefixpatterns}. Specifically, the Node operator is defined by the \textit{fan\_in}, the number of input data, and the \textit{fan\_out}, the number of output data. The radix $r$, which is given by the algorithm pattern, has a direct bearing on the number of steps taken, $K$. Thus, $r$ and $n$ usually appears in the expression that calculates $K$. In general, most of the parallel prefix algorithms use binary Node operators and employ $r=2$; hence, in most cases $N=2^n$. As a set of representative parallel prefix workloads, we have chosen: the scan primitive, the Fast Fourier Transform (FFT), and tridiagonal systems solvers.

In order to apply any tuning methodology, kernels must expose their performance parameters to be tuned. We have chosen the BPLG \footnote{BPLG Library is available at http://bplg.des.udc.es/BPLib.zip} library \cite{BPLGJacobo}, which implements the selected parallel prefix operations. This tuning library is composed of  a set of CUDA skeletons,  also called \textit{building blocks}, written as templates. This allows the user customize them with user-defined datatype and performance parameters. The tunable performance parameters by the CUDA skeletons are the number of \textit{fan\_in} elements per Node operator, the number of Node operators performed per thread, the threadblock size and the amount of shared memory per threadblock. Their optimal values vary for each problem size, algorithm, and target architecture. In addition to these four parameters, some skeletons allow the user to make other choices, for example selecting working on registers (shuffle instructions) instead of shared memory, or declaring the offset between the Node operators in the prefix circuit to be loaded by each thread. Composing this CUDA skeletons enables the solution of the following parallel-prefix operations:

\begin{figure}[t!]
\centering
\subfloat[ Cyclic Reduction pattern.]{\includegraphics[width=0.37\columnwidth]{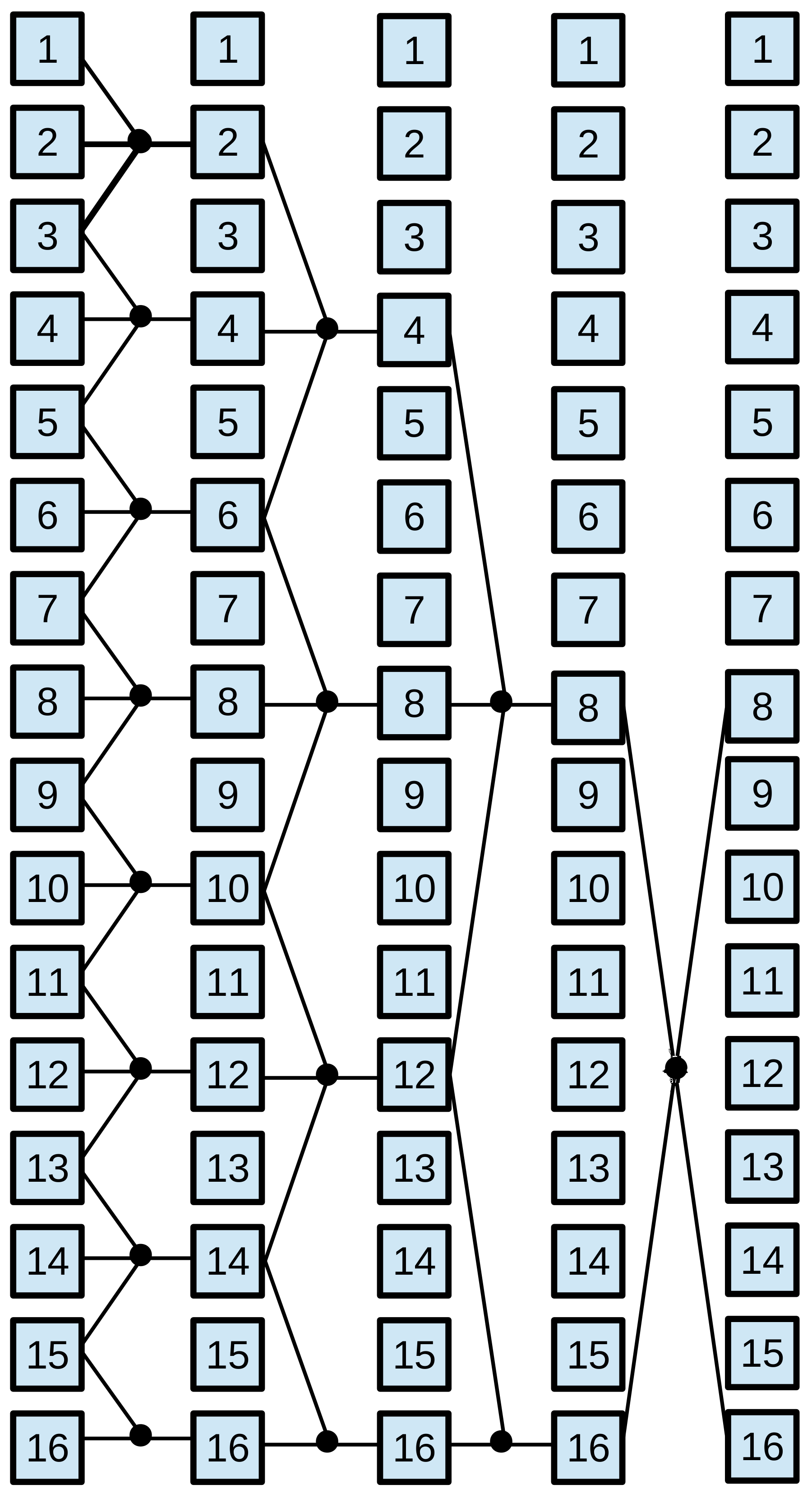}} \qquad
\subfloat[ Parallel Cyclic Reduction pattern.]{\includegraphics[width=0.38\columnwidth]{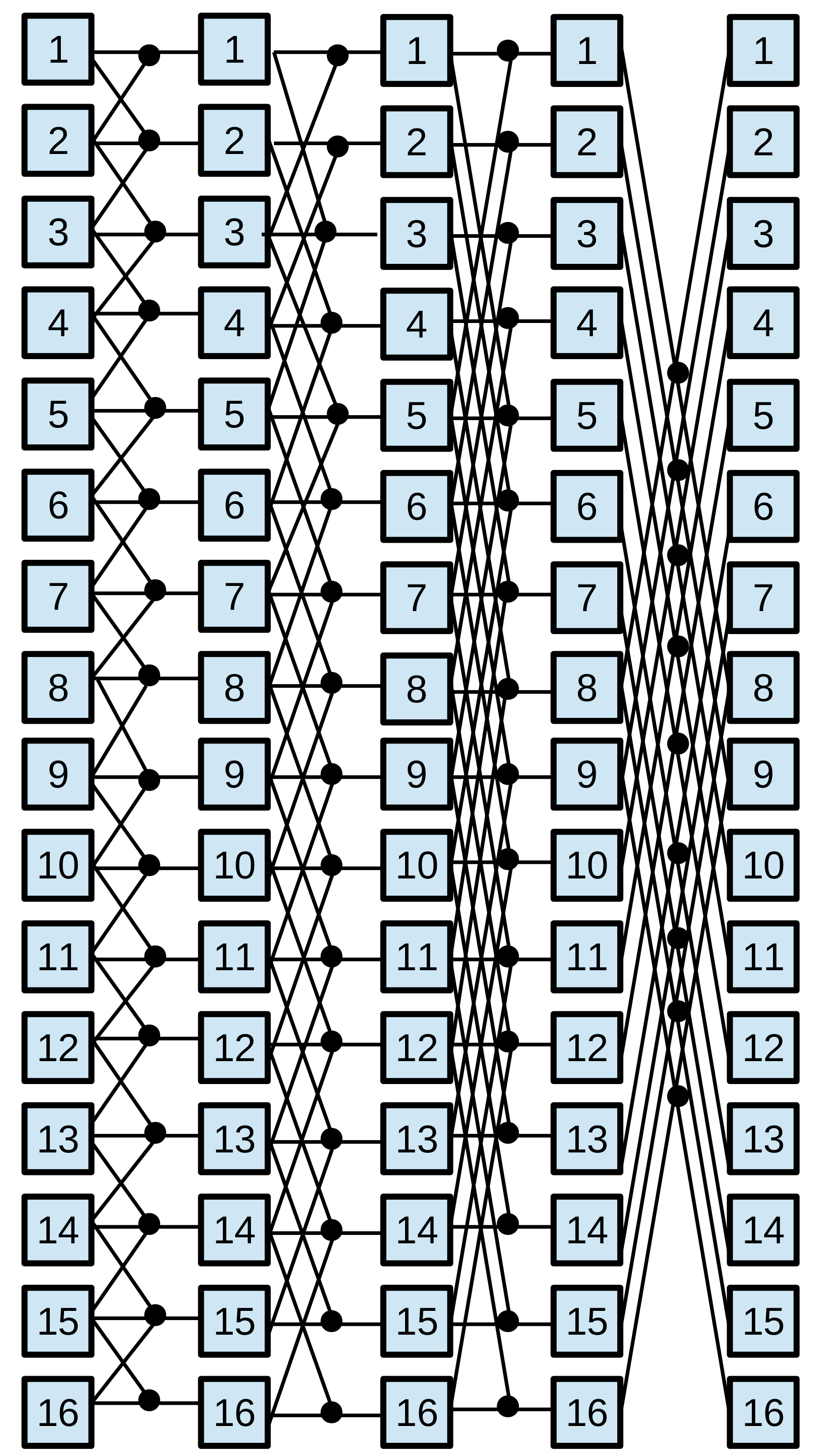}}

\subfloat[Stockham pattern.]{\includegraphics[width=0.52\columnwidth]{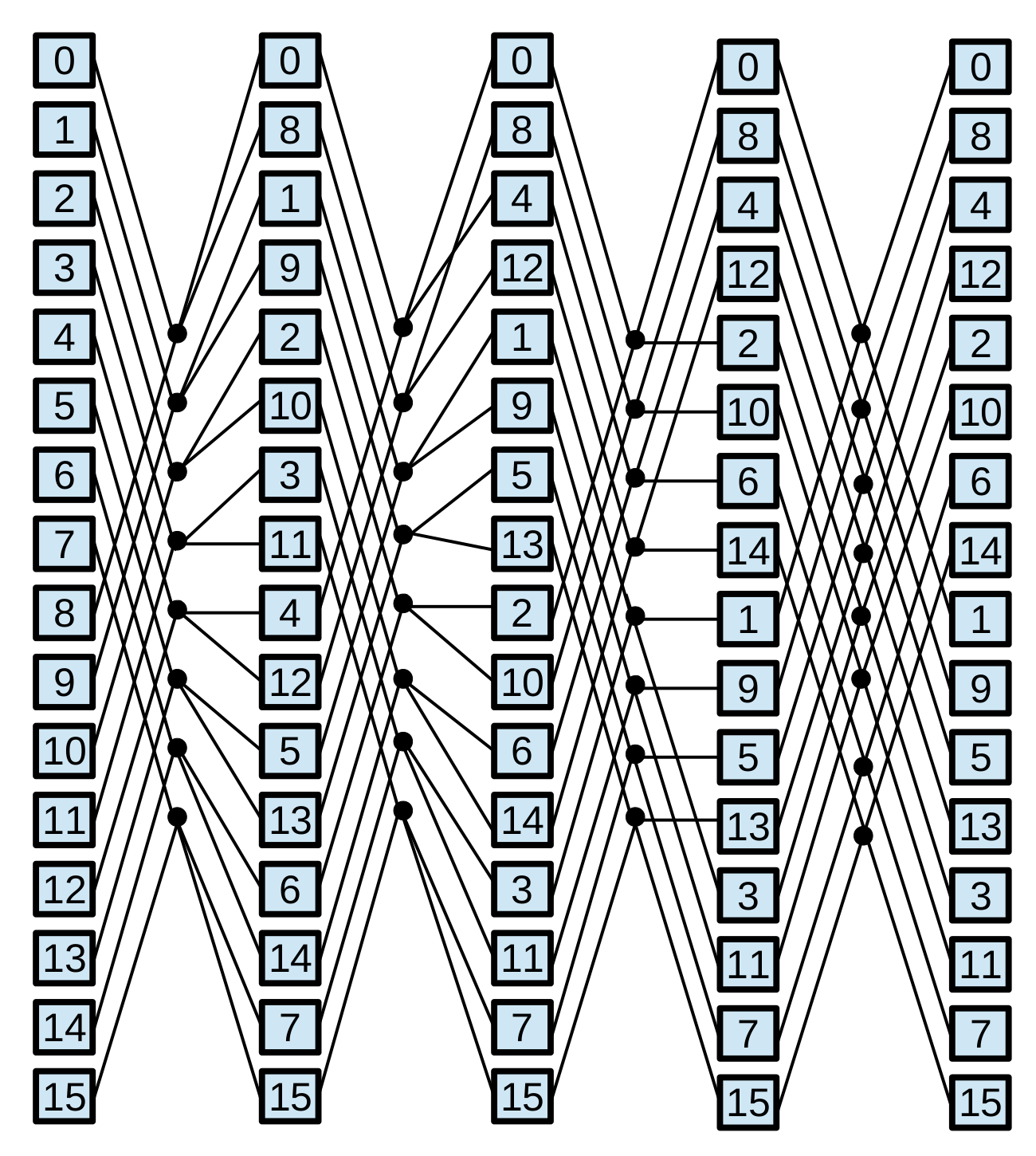}} \qquad
\subfloat[Ladner-Fischer pattern.]{\includegraphics[width=0.35\columnwidth]{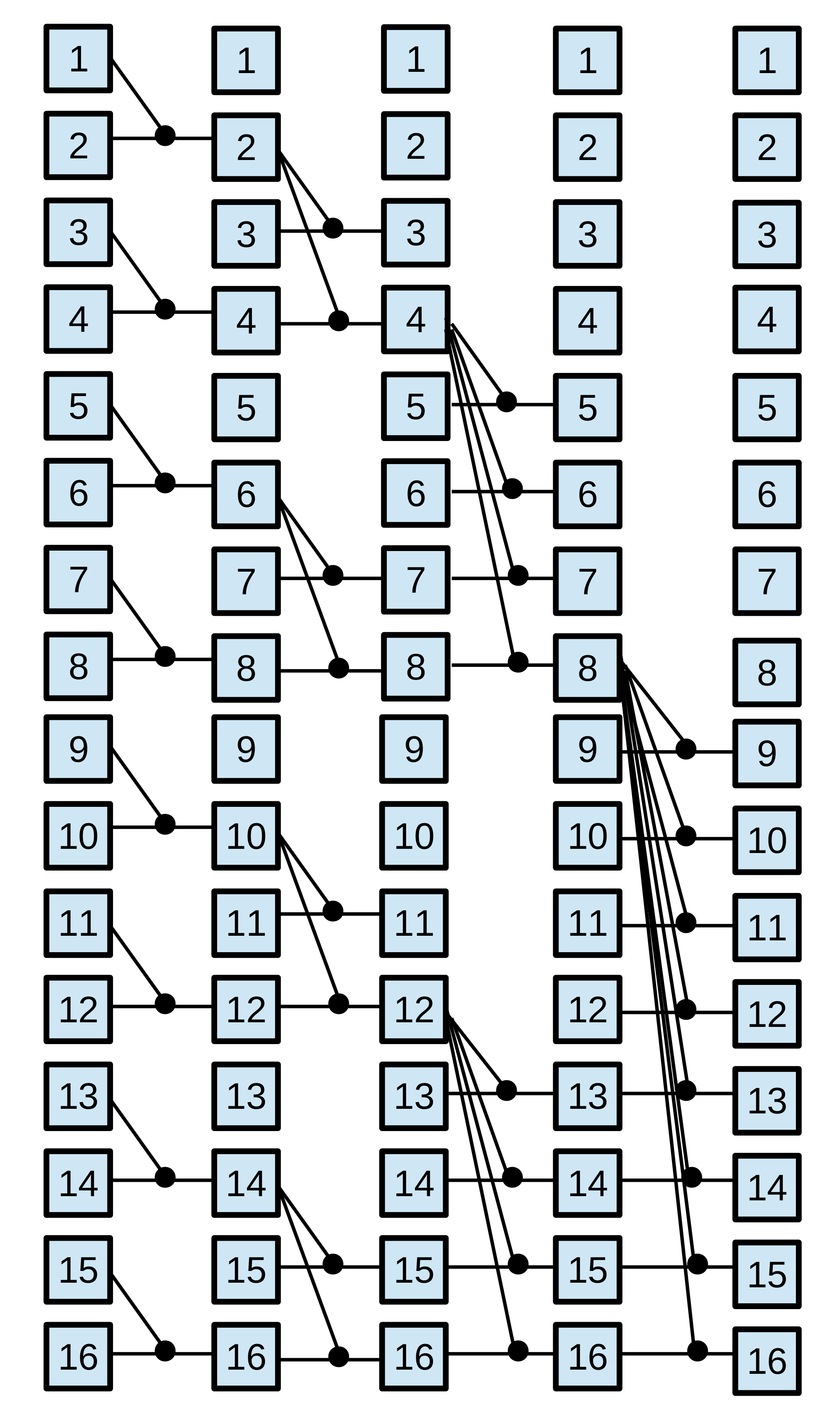}} 

\subfloat[Wang\&Mou pattern.]{\includegraphics[width=0.49\columnwidth]{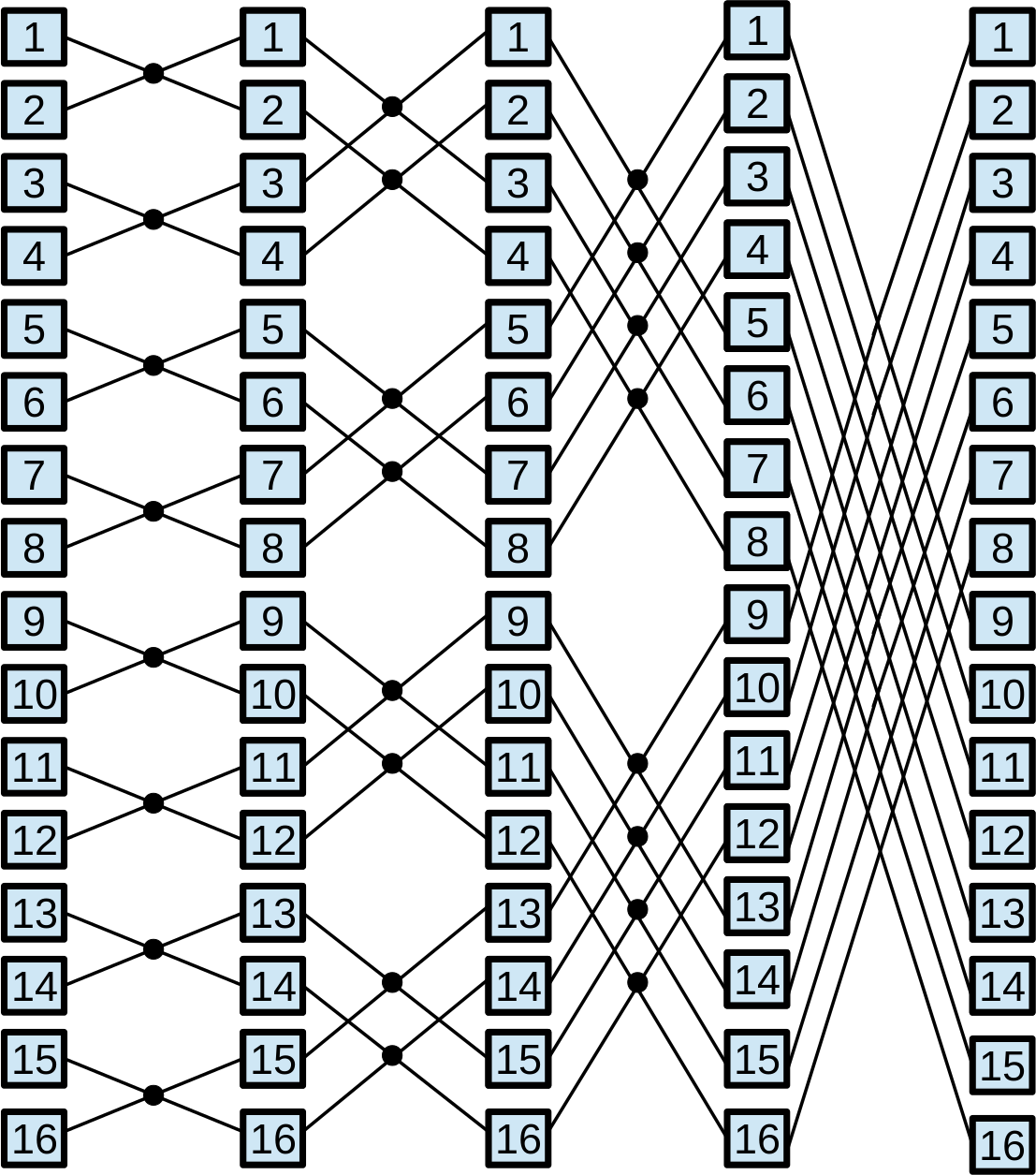}} \qquad
\subfloat[Kogge-Stone pattern.]{\includegraphics[width=0.37\columnwidth]{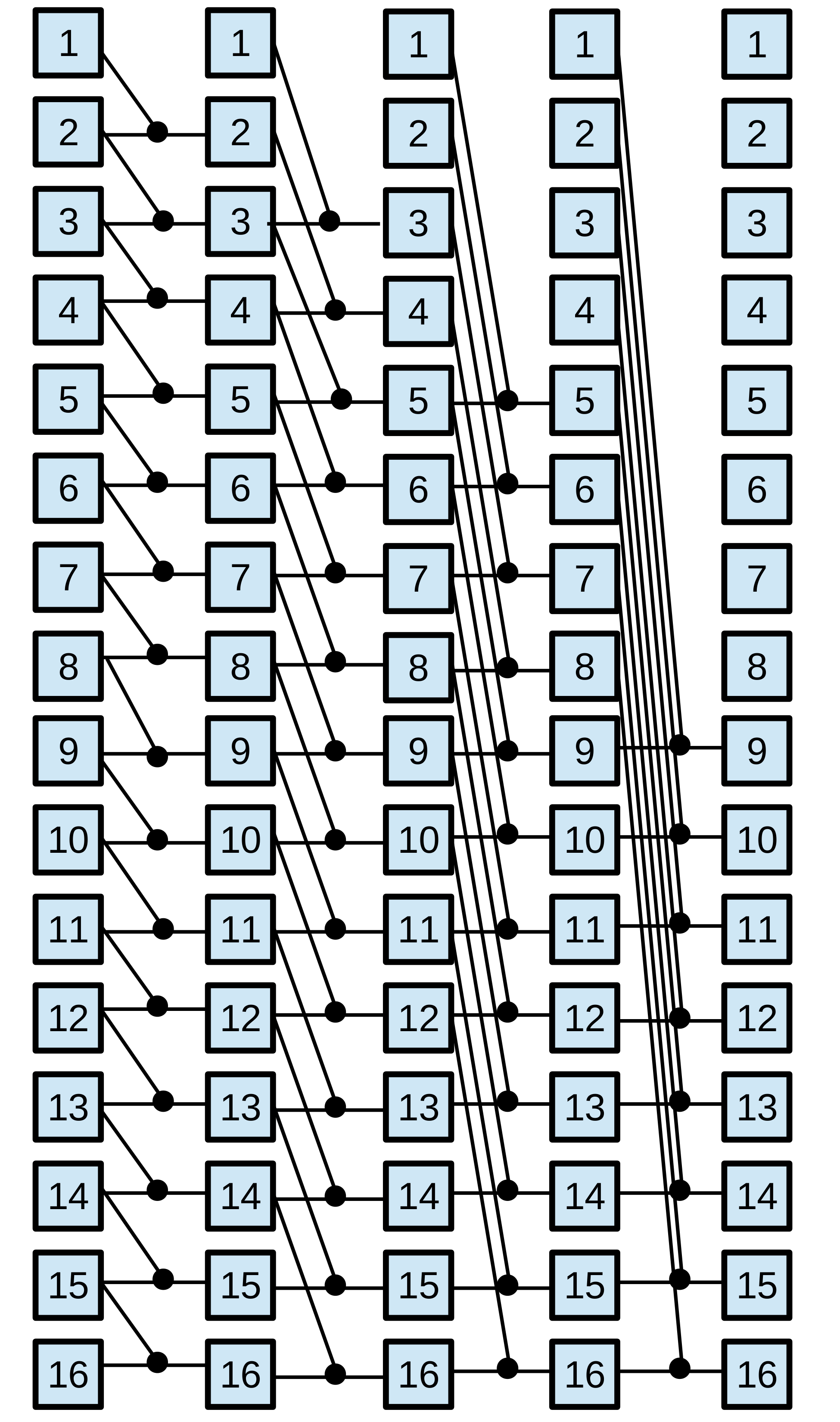}} 
\caption{Different prefix circuits for $N=16$.}
\label{fig:parallelprefixpatterns}
\vspace{-4mm}
\end{figure}

\begin{itemize}
    \item \textbf{Tridiagonal system solvers}. A \textit{tridiagonal system} is composed of \textit{N} equations $E_i$: $a_i x_{i-1}+b_i x_i + c_i x_{i+1} = d_i $. The $b_i$ coefficients constitute the main diagonal of the coefficient matrix, whereas $a_i$ and $c_i$ are known as the lower and upper diagonals, respectively. The BPLG library implements different solvers: Cyclic Reduction (CR)\cite{Hockney:1965:FDS:321250.321259}, Parallel Cyclic Reduction (PCR) \cite{hockney1988parallel}, Ladner-Fischer (LF) \cite{7397622} and Wang\&Mou (WM) algorithm \cite{WANG}, whose patterns are represented in Figure \ref{fig:parallelprefixpatterns}.
    
    \item \textbf{Scan Operator}. This operation replaces each element with the accumulated sum from the first element up to itself. Within the BPLG library, two distinct algorithms can be utilized to perform this computation: a reduction based on the Ladner-Fischer (LF) pattern and another based on the Kogge-Stone (KS) pattern. Figure \ref{fig:parallelprefixpatterns} illustrates both patterns.
    
    \item \textbf{Fast Fourier Transform (FFT)}. BPLG provides a tuning implementation of the complex FFT based on the Stockham algorithm (see Figure \ref{fig:parallelprefixpatterns}), it also allows tuning the radix of the pattern, reducing the number of overall steps.

    \end{itemize}

\section{Tuning Methodologies for Performance Portability on GPU-Embedded Systems}

 Rather than focusing on functional portability, where portable languages can be used among different hardware but not ensuring optimal performance results, this work focuses on finding the optimal performance parameters for each platform, since GPU-embedded systems often execute real-time or time-consuming applications where performance is critical.

 In the context of this paper, it is important to highlight the distinction between online and offline autotuning and how analytical-based and ML-based models perform in these scenarios. Online autotuning refers to the process of dynamically optimizing system parameters during the runtime of an application, while offline autotuning involves optimizing parameters prior to the actual execution of the application. Analytical-based models work well for online autotuning as they provide immediate optimal configurations, especially suitable for real-time applications. On the other hand, ML-based models may introduce an overhead when used for online autotuning, as it requires multiple evaluations to train the surrogate model. However, ML-based models can still be valuable for online autotuning when the time spent on creating the training dataset can be amortized over multiple invocations of the same routine or when the algorithm is iterative, allowing for better performance in subsequent executions.

It is important to emphasize that the problem sizes considered in these methodologies are limited to those supported by the BPLG implementation, specifically small and medium-sized problems that can fit within the CUDA shared memory. As previously mentioned, these methodologies are applicable to any NVIDIA Jetson architecture.

\subsection{An Analytical Model-driven Tuning Methodology for the NVIDIA Jetson}
\label{tuning}

In \cite{jos2018}, BPLG authors presented an analytical tuning methodology for small-medium parallel-prefix problems on CUDA desktop and server GPUs, while our work extends this methodology to tune embedded GPUs. When considering performance modeling for GPU-embedded systems, several factors come into play, such as the limited number of SMs, reduced memory bandwidth, and constrained power consumption. GPU-embedded systems necessitate a distinct approach to performance modeling.

We adhere to the identical set of performance premises and parameters established in \cite{jos2018}: \textit{(i)} a reduction of high-latency communications; \textit{(ii)} a high warp occupancy to hide latency, also taking into account a high block parallelism rate per SM; and \textit{(iii)} a high granularity of work performed by threads (high instruction level parallelism), being aware of register consumption. The attainment of these premises can be accomplished by specifying certain values to the set of performance parameters exposed within the BPLG library: $S$, the number of elements stored in shared memory per threadblock;  $L$, which is the number of threads per threadblock; and $P$ which is the number of elements processed by each thread. They are also related by $S=P\times L$ (except when shuffle operations can replace the communication pattern based on shared memory). These three parameters are represented by the tuple $(S,P,L)$ and their values depend on both the algorithm, the target architecture and $N$. Also, this methodology concentrates on batch execution to effectively utilize all available GPU resources, where each invocation to the library simultaneously executes $G$ batches of size $N=r^n$, $r$ power of two.

%For the first premise, coalescing patterns are essential for achieving the maximum memory bandwidth, especially in memory-bound problems. BPLG already provides CUDA skeletons where the load/store memory operations work on coalescing patterns, but the assignment of Node operators to individual threads can play an important role. Loading neighboring Node operators can significantly influence performance when compared to loading Node operators in a strided manner.

The application of the existing performance premises can be summarized in the following manner. The first premise looks for implementing coalescing patterns to achieve the maximum memory bandwidth. This is crucial in embedded GPUs where the memory bandwidth is reduced. Threads loading Node operators in a strided manner can help to achieve this. Regarding the second premise, GPU parallelism can be determined in terms of the number of threadblocks per SM (\textit{SM block parallelism)}, or by the number of warps per SM (\textit{SM warp parallelism}). A trade-off must be sought between them depending on the problem features. The number of warps per SM, $W_{SM}$, is limited by the maximum number of warps per SM ($W^{max}$ architectural defined) and depends on the number of active threadblocks ($B_a$) being simultaneously executed per SM. Also, $B_a$ is bounded by the architecture design ($B^{max}$) and is constrained by the registers and shared memory available in the \textit{SM}. With respect to the third premise, one thread is responsible for computing at least one Node operator from the parallel prefix circuit, where $P=max(fan\_{in},fan\_{out})$. It may be interesting to process more Node operators per thread, i.e. increase $P$, if extra register consumption does not affect the SM occupancy. This would reduce the number of required threads to compute the same amount of work. We could also increase the radix $r$ of the algorithm, when the communication pattern allows it. Each thread would continue working with a single Node operator, but its \textit{fan\_in} and \textit{fan\_out} values are increased, and consequently $P$ is also increased. This would change the definition of the algorithm owing to $N=r^n$, decreasing the number of steps taken.

The original methodology for desktop and server GPUs in \cite{jos2018} looks for maximizing both SM block and warp parallelism, and balancing them in the cases where is not possible to achieve their maximum values. 
Nevertheless, keeping higher warp occupancy is more important in CUDA embedded GPUs, compared to desktop or server GPUs due to several reasons. Higher warp occupancy reduces the number of idle resources and helps maximize the utilization of available hardware, resulting in more efficient computation with lower power consumption and reduced thermal output, leading to longer battery life. Additionally, many embedded applications, such as robotics, autonomous vehicles, and drones, have strict real-time requirements and low-latency demands. Higher warp occupancy helps ensure more consistent and predictable execution times, minimizing variations in thread execution and reducing overall latency. Prioritizing having a higher warp occupancy can yield greater benefits in GPU-embedded systems compared to having a high number of active thread blocks per SM due to high memory access latency. By keeping a higher warp occupancy, GPU performance can be maximized by effectively hiding memory latency and minimizing the impact of memory stalls, leading to improved overall throughput and resource utilization.
Once a certain level of warp occupancy is achieved, hiding latency is not the main issue anymore. After that, focusing on increasing instruction level parallelism (ILP) can be particularly important for achieving optimal performance compared to server GPUs where resources are relatively more abundant. Thereby, increasing the radix, even when slightly reducing $B_a$, it reduces the number of taken steps (consequently synchronization points) and increases ILP. 

Following this idea, this work proposes a performance guideline that we have found very accurate based on  extensive empirical observations and experience with these devices. Specifically, our tuning guideline for GPU-embedded systems can be summarized as:

\begin{itemize}
  \item Choose $(S,L,P)$ values that achieve both $W_{SM}=W^{max}$ and $B_a=B^{max}$. If this is not possible:
  \item Find the tuple that maximizes $B^a$, while the ratio $W_{SM}/W^{max}$ keeps between $[60\%-100\%]$. If this warp occupancy ratio cannot be reached:
  \item Maximize the warp occupancy. If there are several possibilities, take the one that maximizes $P$.
  \item In all previous steps, if the algorithm allows increasing its radix $r$, then select the configuration that increases $r$ even when reducing $B_a$.
  \end{itemize}

\begin{figure}
\centering
        \subfloat[Parameters that maximize the number of warps and blocks per SM in GM20B.]{
        \resizebox{4.9cm}{!} {
        \begin{tabular}{ccccc}
        \hline
         \begin{tabular}{@{}c@{}} Warps \\ per \\ block\end{tabular} &\begin{tabular}{@{}c@{}} Regs\\ per \\ thread \end{tabular} &  \begin{tabular}{@{}c@{}} Shared \\ memory  \\ per block \end{tabular} &  \begin{tabular}{@{}c@{}} Warp \\ occup.  \end{tabular} &  \begin{tabular}{@{}c@{}}SM \\  blocks  \end{tabular} \\
        \hline
        \hline
       
         1&64&2048&50\%&32 \\
        2&40&0&75\%&24\\
        \hline
        \textbf{2}&\textbf{32}&\textbf{2048}&\textbf{100\%}&\textbf{32}\\
        \hline
        2&40&2560&63\%&20\\
        4&32&4096&100\%&16\\
        4&40&5120&75\%&12\\
        8&32&8192&100\%&8\\
        8&40&10240&75\%&6\\
        16&32&16384&100\%&4\\
        32&32&32768&100\%&2\\
        \hline
        \end{tabular}
        }
        }
    \quad
    \subfloat[TS performance parameters.]{
    \resizebox{3.1cm}{!} {
	\begin{tabular}{c c}
	\hline
	\multicolumn{1}{|c|}{\textit{Size}}&\multicolumn{1}{c|}{  \textit{(S,P,L) values}}\\
	\hline
	\hline
	\multicolumn{2}{|c|}{CR, PCR, LF}\\
	\hline
	\multicolumn{1}{|c|}{$N \leq 64$ }&\multicolumn{1}{c|}{$(0,2,64)$} \\
	\hline
	\multicolumn{1}{|c|}{$N=128$ }& \multicolumn{1}{c|}{$(0,4,64)$}\\
	\hline
	\multicolumn{1}{|c|}{$N>128$ }& \multicolumn{1}{c|}{$(N,2,N/2)$}\\
	\hline	
    \hline
    \multicolumn{2}{|c|}{WM}\\
    \hline
    \multicolumn{1}{|c|}{$N \leq 128$ }&\multicolumn{1}{c|}{$(0,4,64)$} \\
	\hline
	\multicolumn{1}{|c|}{$N>128$ }& \multicolumn{1}{c|}{$(N,4,N/4)$}\\
	\hline
	\end{tabular}
 }
	}
    
    \subfloat[Scan performance parameters.]{
    \resizebox{4cm}{!} {
  \begin{tabular}{c c}
	\hline
	\multicolumn{1}{|c|}{\textit{Size}}&\multicolumn{1}{c|}{\textit{(S,P,L) values}}\\
	\hline
	\hline
	\multicolumn{2}{|c|}{LF pattern}\\
	\hline
	\multicolumn{1}{|c|}{$N \leq 256$ }&\multicolumn{1}{c|}{$(8192/N,4,64)$} \\
	\hline
	\multicolumn{1}{|c|}{$N>256$ }& \multicolumn{1}{c|}{$(32, 4, N/4)$}\\
	\hline	
	\hline
	\multicolumn{2}{|c|}{KS pattern}\\
	\hline
	\multicolumn{1}{|c|}{$N \leq 256$ }&\multicolumn{1}{c|}{$(8192/N,4,64)$} \\
	\hline
	\multicolumn{1}{|c|}{$N>256$ }& \multicolumn{1}{c|}{$(32, 4, N/4)$}\\
	\hline	
	\end{tabular}	
    }
    }
    \quad
    \subfloat[FFT performance parameters]{
    \resizebox{4cm}{!} {
    \begin{tabular}{c c}
	\hline
	\multicolumn{1}{|c|}{\textit{Size}}&\multicolumn{1}{c|}{  \textit{(S,P,L) values}}\\
	\hline
	\hline
	\multicolumn{1}{|c|}{$N \leq 256$ }&\multicolumn{1}{c|}{$(256,4,64)$} \\
	\hline
	\multicolumn{1}{|c|}{$N>256$ }& \multicolumn{1}{c|}{$(N,4,N/4)$}\\
	\hline
	\end{tabular}
    }
    }
    \caption{Performance Parameters for the parallel-prefix operations using the analytical-based search. The highlighted row represents the configuration that maximizes both warp  and threadblock occupancy.}
	\label{analytical-tuning}	
 \end{figure}

 Considering the GM20B GPU and proposed guideline, it is necessary to find the optimal values of the $(S,P,L)$ parameters for each algorithm. Figure \ref{analytical-tuning} (a) summarizes the parallelism and the resource consumption achieved by different threadblock configurations. \\

\subsection{A ML-based Tuning Methodology for the NVIDIA Jetson}

%This work implements the tuning methodology proposed in \cite{}, based on Bayesian optimization and transfer learning, to be used for the NVIDIA Jetson TX1. This methodology defines three sets of ..... and Platform Parameters, $\mathcal{C}$, which represent the underlying architecture in terms of hardware features.  Given an input $a_j\in\mathcal{A}$ and platform $c_k\in\mathcal{C}$, the objective is to search for the optimal vector of performance parameters $x_B\in\mathcal{B}$, defined as: \argmin_{x_B \in \mathcal{B}} f(X) : X=(a_j,c_k,x_B) 

We use two sets of parameters participating in the tuning search \cite{cox}:  Input Parameters, $\mathcal{A}$, which characterize the input in terms of computational shapes or data layout of the application; and Performance Parameters, $\mathcal{B}$, the parameters to be optimized for the target architecture.  Given an input $a_j\in\mathcal{A}$, our goal is to search for the optimal vector of performance parameters $x_B\in\mathcal{B}$ that meets $\argmin_{x_B \in \mathcal{B}} f(X) : X=(a_j,x_B)$. Here, $f(\cdot,\cdot)$ denotes the objective function that maps each configuration to a corresponding execution time. Calculating derivatives for $f$ is indoable \cite{Dorier_2022}, necessitating actual evaluations of different configurations. In this methodology, a Bayesian optimization search is employed, as it has been shown to be effective in exploring promising regions of the parameter space \cite{BO}, thereby minimizing the number of required evaluations. The procedural workflow for Bayesian optimization is outlined as follows: First, a small set of configurations are randomly sampled from the search space and evaluated. The resulting data (configurations and execution times) are added to a dataset that trains the surrogate model. Based on the actual predictions of the surrogate model, the acquisition function optimizes a score, guiding the selection of the next configuration to be evaluated. The iterative process continues until a stopping criteria is met. To do this Bayesian-optimization search, we use the GPTune framework \cite{gptune} that uses the \textit{Linear Coregionalization Model} (LCM) \cite{gptune13} as surrogate model, and the Expected Improvement acquisition function \cite{BO34}. % As demonstrated in \cite{}, the number of samples required to train the surrogate model can be reduced if the training is accelerated with transfer learning by using performance data from other platforms for the same problem. If the source and the target portability scenarios are similar, LCM will learn significant correlations among them during the training, accelerating the search process. To apply the cross-platform transfer learning, which is not supported by the framework, the platform parameters are embedded as input parameters and some metadata is modified to enable the use of data from another platform.

As a summary of the process, the first step is to define the corresponding Input Parameters, $\mathcal{A}$, and Performance Parameters, $\mathcal{B}$, together with any constraint in order to define the search space. Then, this search space is explored with the searching workflow, which orchestrates the search by calling the Bayesian-optimization framework. The interactive search is run for each algorithm on the target platform until a user-defined stop criteria is met. For our BPLG parallel-prefix search, the Input and Performance parameters are described in Table \ref{parameters-TL}. The problem size describes the input problem we are solving, and its range of values depends on the BPLG implementation for each operation. Regarding the performance parameters, the range of taken values, which defines the search space, also depends on the parallel-prefix algorithm implementation being tuned, as it is seen in the next section. We also added \textit{shuffle} as a binary performance parameter that chooses an alternative communication based on warp shuffles instead of shared memory, when available. With respect to the stopping criteria, we perform a sliding-window check and stop the search when no progress has happened in the last iterations, which saves extra evaluations once a good optimum has been found. In this search, we stop the search if no progress within the last 5 evaluations. Also, we set a high execution-time value for those executions with configurations that are invalid or are not finishing after 1 minute.

\begin{table}
	\centering
	%\resizebox{6cm}{!} {
	\begin{tabular}{c c}
	\multicolumn{2}{l}{\textbf{Input Parameters}}\\
	\hline
	\multicolumn{1}{|c|}{\textit{Problem Size}}&\multicolumn{1}{c|}{ \parbox{6cm}{Any power of two allowed by the implementation.}} \\
    \hline
	\multicolumn{2}{l}{\textbf{Performance Parameters}}\\
	\hline
	\multicolumn{1}{|c|}{$S$} &\multicolumn{1}{c|}{ \parbox{6cm}{Elements stored in Shared Memory. A power of two.}}\\
	\hline
	\multicolumn{1}{|c|}{$P$} &\multicolumn{1}{c|}{ \parbox{6cm}{Number of elements processed by thread. A power of two.}}\\
	\hline
	\multicolumn{1}{|c|}{$L$ }&\multicolumn{1}{c|}{\parbox{6cm}{ Number of threads per threadblock. Power of two limited by $max\_threads\_block$.}}\\
 	\hline
	\multicolumn{1}{|c|}{$r$ }&\multicolumn{1}{c|}{\parbox{6cm}{ The applied \textit{radix} in the prefix pattern. Power of two. }}\\
	\hline 
    \multicolumn{1}{|c|}{$shuffle$ }&\multicolumn{1}{c|}{\parbox{6cm}{ If allowed, the use of shuffle communications instead of Shared Memory.}}\\
	\hline 
	\end{tabular}	
	%}
	\caption{ML-search: Input and Performance Parameters for the BPLG library.}
	\label{parameters-TL}	
 %\vspace{-4mm}
	\end{table}

\subsection{Tuning Larger Problem Sizes}

Both methodologies can be extended to larger problem sizes. When the problem size exceeds the capacity of shared memory ($N>S$), collaborative computation among multiple thread blocks becomes necessary. Two options are available in recent CUDA versions to synchronize these thread blocks: launching multiple kernels using a multi-kernel strategy or utilizing Cooperative Groups (available from CUDA 9.0 onwards). However, the number of thread blocks that can participate in Cooperative Groups is limited by the maximum number of resident thread blocks per Streaming Multiprocessor (SM). The resource consumption of BPLG kernels can significantly reduce the number of resident thread blocks per SM, resulting in fewer threadblocks than the number required to complete the collaborative computation. BPLG implements FFT for problem sizes exceeding shared memory capacity, and our work aims to tune this implementation as well.

In their work \cite{7970194}, the authors of BPLG presented an analytical model for tuning BPLG large-sized parallel-prefix operations on desktop and server GPUs, employing a multi-kernel strategy. Building on this idea, we extend our analytical heuristics for small to medium sizes on CUDA-embedded systems with the premise that  \textit{The number of kernels $m$ needs to be minimized}. If $N=r^n$ and $S=r^s$, then the number of required kernels $m$ can be calculated as $m=\lceil \frac{n}{s} \rceil$. Once the optimal $S$ is determined using this expression, our previous guideline for small to medium problems is applied to tune the $(S,P,L)$ and radix $r$ values for each resulting kernel. The inclusion of multiple kernels has introduced additional complexities in the tuning procedure, as the $(S,P,L)_m$ values are interdependent. The value of these parameters in one kernel impacts the workload of others. 

 It is noteworthy to mention that the ML-based methodology will treat the tuning process as a black box. We just have to extend the parameter space, but the underlying intricacies and interdependencies among kernels are transparent to this purpose.

\section{Finding the BPLG performance values with our methodologies}

\begin{figure}
\centering
        \subfloat[TS performance parameters.]{
    \resizebox{4cm}{!} {
	\begin{tabular}{|c| c| c|}
	\hline
	\begin{tabular}{@{}c@{}} $N$ \end{tabular}&\begin{tabular}{@{}c@{}} Required \\ evaluations \end{tabular} & \begin{tabular}{@{}c@{}} $(S,P,L)$ \\ values \end{tabular}\\
	\hline
	\hline
	\multicolumn{3}{|c|}{WM}\\
	\hline
	\multicolumn{1}{|c|}{$64$ }&\multicolumn{1}{|c|}{20}&\multicolumn{1}{c|}{$(0,4,128)$} \\
	\hline
	\multicolumn{1}{|c|}{$128$ }&\multicolumn{1}{|c|}{15}& \multicolumn{1}{c|}{$(0,4,128)$}\\
	\hline
	\multicolumn{1}{|c|}{$256$ }&\multicolumn{1}{|c|}{15}& \multicolumn{1}{c|}{$(512,4,128)$}\\
	\hline	
    \multicolumn{1}{|c|}{$512$ }&\multicolumn{1}{|c|}{5}& \multicolumn{1}{c|}{$(512,2,256)$}\\
	\hline
    \multicolumn{1}{|c|}{$1024$ }&\multicolumn{1}{|c|}{5}& \multicolumn{1}{c|}{$(1024,4,256)$}\\
	\hline
	\end{tabular}
    }
	}
    \quad
    \subfloat[Scan performance parameters.]{
    \resizebox{4cm}{!} {
	\begin{tabular}{|c| c| c|}
	\hline
	\begin{tabular}{@{}c@{}} $N$ \end{tabular}&\begin{tabular}{@{}c@{}} Required \\ evaluations \end{tabular} & \begin{tabular}{@{}c@{}} $(S,P,L)$ \\ values \end{tabular}\\
	\hline
	\hline
	\multicolumn{3}{|c|}{LF-scan}\\
	\hline
	\multicolumn{1}{|c|}{$64$ }&\multicolumn{1}{|c|}{10}&\multicolumn{1}{c|}{$(128,2,128)$} \\
	\hline
	\multicolumn{1}{|c|}{$128$ }&\multicolumn{1}{|c|}{15}& \multicolumn{1}{c|}{$(128,4,128)$}\\
	\hline
	\multicolumn{1}{|c|}{$256$ }&\multicolumn{1}{|c|}{10}& \multicolumn{1}{c|}{$(32,4,64)$}\\
	\hline	
    \multicolumn{1}{|c|}{$512$ }&\multicolumn{1}{|c|}{10}& \multicolumn{1}{c|}{$(64,4,256)$}\\
	\hline
    \multicolumn{1}{|c|}{$1024$ }&\multicolumn{1}{|c|}{5}& \multicolumn{1}{c|}{$(32,2,512)$}\\
	\hline
    \multicolumn{1}{|c|}{$2048$ }&\multicolumn{1}{|c|}{5}& \multicolumn{1}{c|}{$(32,4,512)$}\\
	\hline
    \multicolumn{1}{|c|}{$4096$ }&\multicolumn{1}{|c|}{5}& \multicolumn{1}{c|}{$(32,4,1024)$}\\
	\hline
	\end{tabular}
    }
	}
    
    \subfloat[FFT performance parameters.]{
    \resizebox{4cm}{!} {
	\begin{tabular}{|c| c| c|}
	\hline
	\begin{tabular}{@{}c@{}} $N$ \end{tabular}&\begin{tabular}{@{}c@{}} Required \\ evaluations \end{tabular} & \begin{tabular}{@{}c@{}} $(S,P,L)$ \\ values \end{tabular}\\
	\hline
	\hline
	\multicolumn{1}{|c|}{$64$ }&\multicolumn{1}{|c|}{10}&\multicolumn{1}{c|}{$(256,8,32)$} \\
	\hline
	\multicolumn{1}{|c|}{$128$ }&\multicolumn{1}{|c|}{10}& \multicolumn{1}{c|}{$(512,8,64)$}\\
	\hline
	\multicolumn{1}{|c|}{$256$ }&\multicolumn{1}{|c|}{10}& \multicolumn{1}{c|}{$(256,4,64)$}\\
	\hline	
    \multicolumn{1}{|c|}{$512$ }&\multicolumn{1}{|c|}{10}& \multicolumn{1}{c|}{$(512,2,256)$}\\
	\hline
    \multicolumn{1}{|c|}{$1024$ }&\multicolumn{1}{|c|}{5}& \multicolumn{1}{c|}{$(1024,4,256)$}\\
	\hline
    \multicolumn{1}{|c|}{$2048$ }&\multicolumn{1}{|c|}{5}& \multicolumn{1}{c|}{$(2048,4,512)$}\\
	\hline
    \multicolumn{1}{|c|}{$4096$ }&\multicolumn{1}{|c|}{5}& \multicolumn{1}{c|}{$(4096,4,1024)$}\\
	\hline
	\end{tabular}
    }
	}
    \quad
    \subfloat[ Required evaluations for larger FFT sizes ]{
    \resizebox{4cm}{!} {
	\begin{tabular}{|c| c| c| c|}
	\hline
	\begin{tabular}{@{}c@{}} $N$ \end{tabular}&\begin{tabular}{@{}c@{}} Eval. \end{tabular} & \begin{tabular}{@{}c@{}} $N$ \end{tabular} &\begin{tabular}{@{}c@{}} Eval. \end{tabular}\\
	\hline
	\hline
	\multicolumn{1}{|c|}{$2048$ }&\multicolumn{1}{|c|}{5}&\multicolumn{1}{|c|}{$262144$ }&\multicolumn{1}{|c|}{15}\\
	\hline
	\multicolumn{1}{|c|}{$4096$ }&\multicolumn{1}{|c|}{5}&\multicolumn{1}{|c|}{$524288$ }&\multicolumn{1}{|c|}{25} \\
	\hline
	\multicolumn{1}{|c|}{$8192$ }&\multicolumn{1}{|c|}{15}&\multicolumn{1}{|c|}{$1048576$ }&\multicolumn{1}{|c|}{40} \\
	\hline	
    \multicolumn{1}{|c|}{$16384$ }&\multicolumn{1}{|c|}{10}&\multicolumn{1}{|c|}{$2097152$ }&\multicolumn{1}{|c|}{35} \\
	\hline
    \multicolumn{1}{|c|}{$32768$ }&\multicolumn{1}{|c|}{15}&\multicolumn{1}{|c|}{$4194304$ }&\multicolumn{1}{|c|}{30} \\
	\hline
    \multicolumn{1}{|c|}{$65536$ }&\multicolumn{1}{|c|}{20}&\multicolumn{1}{|c|}{$8388608$ }&\multicolumn{1}{|c|}{40} \\
	\hline
    \multicolumn{1}{|c|}{$131072$ }&\multicolumn{1}{|c|}{20}&\multicolumn{1}{|c|}{ }&\multicolumn{1}{|c|}{} \\
	\hline
	\end{tabular}
    }
	}
    \caption{Performance Parameters and required number of candidate evaluations for the parallel-prefix operations using the ML-based search}
	\label{analytical-tuning-ML}	
 \end{figure}

This section analyzes the resulting performance parameters for each parallel-prefix operation when our methodologies are applied. In all cases, data are in single precision. It is important to note that, due to space constraints, we have only presented the performance parameters obtained through the ML-based search for a subset of the algorithms. In Section \ref{experimental}, we consider all recommended performance configurations provided by the ML-based search when computing performance metrics.

\subsection{Tridiagonal System Solvers}
\label{tstuningsection}

 Firstly, it should be pointed out that each element is an equation composed of 4 simple-precision coefficients, and only WM allows modifying the radix definition. With respect to the analytical-based tuning, the optimal configuration which maximizes both block and warp parallelism, following Figure \ref{analytical-tuning}(a), is achieved with $L=64$ and fewer than 32 registers per thread (the 3rd row in Table). Starting with the CR pattern and taking into consideration additional variables and index calculation, $P$ must be equal to 2 in order to consume less than 32 registers per thread. When $N\leq 64$, each problem is solved with 32 threads and $P=2$ elements per thread. Thus, shuffle instructions can be employed within the warp instead of shared memory. Please observe that each threadblock is executed with $L=64$ threads, thus 2 batches are simultaneously solved. When $N=128$, there are two alternatives: either using $P=4$, 32 threads per problem (2 problems per threadblock) and shuffle communications; or $P=2$, with 64 threads per problem (one problem per threadblock) and the use of shared memory to perform the communications ($128 \ el. \ \times 4 \ coef. \ \times 4 \ Bytes= 2048 \ Bytes$). In the first case, 63\% of warp parallelism and 20 threadblocks are achieved (4th row in Figure \ref{analytical-tuning}(a)); whereas the second case obtains  100\% of warp occupancy, but only 16 threadblocks (5th row in Table) and less bandwidth in communications. Following our performance recipe, the first alternative is chosen. For the remaining problem sizes, ($N>128$), shared memory is needed for storing the problem data; thus $S=N$. When $S$ occupies more than 3072 bytes, the case of $N>128$, it is not possible either the maximum warp parallelism or the maximum block parallelism, choosing the configurations that meet the tuning strategy requirements. Due to the high register consumption, $P=2$ is established for these problem sizes. The same tuning values are obtained for PCR. In the case of LF, each element is composed of two equations ($2 \times 4 \times 4$ bytes per element). As the number of registers must not exceed 32, considering registers employed for additional variables, $P$ must be strictly 2, but the same performance parameters are obtained despite of this restriction following Figure \ref{analytical-tuning}(a). Finally, WM allows modifying the radix of the algorithm due to its regular communication pattern. Hence, in the case of having several reasonable configuration alternatives, increasing $P$ must be prioritized. Specifically, when $N\leq 128$, $(S,P,L)=(0,4,64)$ is used, achieving 75\% warp occupancy and 24 out of 32 threadblocks, performing shuffle communications within each warp. For the remaining sizes,  $(S,P,L)=(N,4,N/4)$ is used.

Regarding the ML-based tuning, the methodology will search for $L$ values between 32 and 1024; $P$ values among $\{2,4,8\}$ (higher values are unrealistic due to high register pressure) and $S$ can take values up to $2048$ elements (32 KB per threadblock). Although it is possible to use up to $48KB$ per threadblock, that would not represent any $N=r^n$ size power of radix. The radix $r$ is set to 2 for CR, PCR and LF, since larger values are not allowed for these patterns, but the model does explore the set $r=\{2,4,8\}$ in WM. The $shuffle$ parameter is only allowed when $N/P \leq 32$, i.e. the computation of each problem can be performed within a warp. Also, the restrictions \textit{(!shuffle OR S==0)} and \textit{(!shuffle AND S=PxL)} have to be met, because the shuffle optimization releases the need of shared memory. Figure \ref{analytical-tuning-ML}(a) shows the suggested configuration for the WM algorithm, together with the number of candidate evaluations required to train the surrogate model. Radix $r$ consistently aligns with the recommended $P$ value, and the  radix $r$ is omitted in the figure. Also, shuffle optimizations were applied whenever $S=0$ in the figure. The higher problem sizes, the fewer configurations are valid, reducing the number of evaluated candidates in the search.

\subsection{Scan Primitive}

BPLG implements the scan primitive with shuffle instructions where 32 elements per problem are stored in shared memory \cite{jos2018}. Examining Table \ref{analytical-tuning}(a), the maximum warp and block parallelism is achieved with $L=64$ and fewer than 32 registers per thread. Considering auxiliary variables and index calculation, $P$ must be less than or equal to 4. Additionally, shared memory consumption per threadblock must be fewer than 2048 bytes; thus, $S\leq 512$ single-precision elements. As $L=64$ and $P=4$,  the number of problems per threadblock can be expressed with $\frac{64}{N/4}=256/N$. Since each problem uses 32 shared memory elements, $S=32\times \frac{256}{N}=8192/N$. When $N>256$, $L$ has to be increased and only one problem can be solved per threadblock, thus $L=N/4$ and $S=32$. The KS pattern obtains the same values by following the same reasoning. Figure \ref{analytical-tuning}(c) summarizes the performance parameters for the LF and KS patterns. 

With respect to the $(S,P,L)$ exploration in the ML-based tuning, the search space is similar to the one seen with tridiagonal solvers. However, the radix of these algorithms cannot be explored, while $P$ can take any value from $P=\{2,4,8,16,32\}$. Due to the given BPLG implementations, $shuffle$ is always set to one, and $S=L/32$. Therefore, the ML-based tuning search only explores values for $P$ and $L$ parameters, and when raising $N$, the amount of possible configurations is highly reduced due to restrictions, leading to a minimal number of candidate evaluations. Figure \ref{analytical-tuning-ML}(b) displays the suggested configurations for the scan LF algorithm.

\subsection{Fast Fourier Transform (FFT)}

The FFT communication pattern allows extending its radix. Thus, increasing $P$ reduces the number of computing steps, internal communications and synchronizations. Referring to Figure \ref{analytical-tuning}(a) as a point of reference once again, $L=64$ should be chosen. Using $P=4$ exceeds the number of 32 registers, but the benefits of having $P=4$ are high in terms of reducing computing steps, as mentioned in Section \ref{tstuningsection}. For $N \leq 256$, there are several problems being solved in each threadblock, as many as $\frac{64 \times 4}{N}$. As $S=P\times L$ and each element is composed of two floats, $S=256$ elements, achieving 75\% of warp parallelism. If $S=4096$, it implies $4096 \times 2 \times 4 bytes= 32KB$ of shared memory per threadblock and leads to only one active threadblock per SM. To avoid it, BPLG first exchanges the real part, performing the computations, and then exchanging the imaginary part, reducing the shared memory to 16KB per threadblock. For the remaining sizes, $L=N/4$ and $S=N$, as Figure \ref{analytical-tuning}(d) shows.

Regarding the ML-based search, radix $r$ can take any value between $r=\{2,4,8,16\}$, and consequently $P=\{2,4,8,16\}$. Also, $shuffle$ is always set to 0, as BPLG does not support shuffle communications for the FFT implementations. Considering $8$ bytes per element and the explained shared-memory multiplexing technique, $S \leq 8192$ elements. Figure \ref{analytical-tuning-ML}(c) shows the achieved configurations for the FFT algorithm with the ML-based search. It always coincides the suggested radix $r$ with the recommended $P$ value, resulting in the omission of the radix $r$ in the Figure.    

\subsection{Larger Problem Sizes: FFT}

We consider problem instances within the range of $4096 < N \leq 8388608$ elements to represent this type of problems. Starting with the analytical-based tuning, $S$ must be equal to 2048 (the maximum number of elements allowed) to minimize the number of required kernels with the multi-kernel strategy. As $2048=P \times L$, if we apply the given performance guideline, then the tuple $(2048,8,512)$ with radix-8 is chosen as maximizes the number of active warps (75\%) per SM over other solutions for this equation. When $N \geq 524288$, three kernels have to be launched.  

For small and medium problem sizes ($N\leq 4096$), the ML-based search could be deemed overkill, as a exhaustive search with few evaluations could have sufficed. However, when the search space grows, the ML-based approach fits better within the study work, offering a coherent and efficient approach to tackle the search. Figure \ref{analytical-tuning-ML}(d) shows the required number of candidate evaluations to train the model. When three kernels must be tuned, the exhaustive search have to explore hundreds of valid combinations, while the ML-based methodology finds a good optimum with few evaluations.

\section{Experimental Results}
\label{experimental}

This section presents the performance results for the cases of study exposed above. Initially, we conduct a comparison of the tuning results obtained from each methodology, followed by a performance comparison of the tuned BPLG implementation against other state-of-the-art libraries. While a direct comparison between the proposed analytical methodology for GPU-embedded systems and the existing methodology for desktop and server GPUs proposed in \cite{jos2018} may be of interest, it was intentionally excluded. We specifically emphasize the comparison between human-developed analytical model-driven methodologies with machine-learning methodologies, as our main goal is to assess effectiveness of two different kind of predictive tuning strategies in GPU-embedded systems rather than benchmarking two specific analytical models.  Also, when comparing the performance of the BPLG implementations to other libraries, the BPLG configurations were tuned to maximize performance, regardless of the methodology employed. Whether using the analytical model-driven approach or the ML-based search, the BPLG implementations were lately fine-tuned with the best configuration found for each problem size.  In all cases, test data are already on the GPU; thus, there are no data transfers during the benchmarks to prevent interactions with other factors in the study. The experiments are run in single precision. Batch execution is used to process $2^{26}/N$ simultaneous problems, therefore as the input size increases, the number of batch executions decreases.

The NVIDIA's Jetson TX1 is used as test platform in this work, which was described in Section \ref{jetsondes}. Specifically, this development board was flashed with the \textit{JetPack 3.3.4 L4T 28.5}, which installed CUDA 9.0 SDK for the purpose of ensuring compatibility with BPLG. In addition to this, due to issues encountered during the installation of GPTune Python dependencies in the device, we opted to run the GPTune Python framework on the host machine and pass the execution candidates to the device. However, it is worth noting that this incompatibility with Python packages has been fixed in newer JetPack and Jetson versions, and the complete Bayesian optimization workflow can be now executed directly on the device. 

Furthermore, the two proposed methodologies cannot guarantee that the suggested optimal configuration is actually the best configuration from the search space. It requires a metric that quantifies the achieved performance with respect to the best possible. Based on the quantitative metric presented in \cite{pennycook}, we proposed using the following metric to capture the performance of a parallel-prefix algorithm $a$ solving a problem size $p$ across all possible problem sizes $C$, given by the harmonic mean:
\begin{align*}
    \Phi (a,C) = 
    \frac{|C|}{\sum_{i \in C }1/e_i(a,p_i) }
\end{align*}
where $e_i(a,p_i)$ is the performance efficiency of algorithm $a$ solving problem size $p_i$, measured as a fraction of the best empirical-observed performance. Thus, an exhaustive search of the parameter space is also executed for each algorithm and problem size to compare the performance efficiency of the suggested configurations. This metric gives the same weight to all problem sizes of the input space. A value of $\Phi=1$ indicates a best match, and lower values measure how much the observations deviate from the best. The BPLG implementation is the responsible in achieving superior performance compared to alternative libraries; however, effective tuning is crucial for BPLG to surpass them. Notably, BPLG exhibits significant performance variations depending on the chosen values for the investigated performance parameters. By analyzing this $\Phi$ metric and comparing performance against other state-of-the-art libraries, we gain valuable insights into the efficacy of our proposed methodologies.

\begin{table}[t!]
	\centering
	\resizebox{9.1cm}{!} {
	\begin{tabular}{c c c c c}
	\hline
	\textit{Solver}&\begin{tabular}{@{}c@{}} Analytical-based \\ Performance \end{tabular}&  \begin{tabular}{@{}c@{}} ML-based \\ Performance \end{tabular}& Analytical $\Phi$ & ML $\Phi$\\
	\hline
	\hline
	BPLG-WM &670 MRows/s&  677 MRows/s&0.9895
&1\\
	\hline
	BPLG-CR &640 MRows/s &635 MRows/s&0.9941&0.9855 \\
	\hline
	BPLG-PCR&547 MRows/s & 545 MRows/s &0.9985&0.9941 \\
	\hline	
	BPLG-LF &668 MRows/s& 669 MRows/s&0.9951&0.9968\\
	\hline	
    \hline
	BPLG-ScanLF &18.27 MData/s& 18.72 MData/s&0.9699& 1\\
	\hline
	BPLG-ScanKS &17.9 MData/s & 17.63 MData/s& 1 &0.9819  \\
    \hline
    \hline
    BPLG-FFT & 39 GFlops/s & 39.09 GFlops&0.9718&0.9813 \\
    \hline
    \hline
    BPLG-Large-FFT & 57.28 GFlops/s & 63.37 GFlops & 0.8739 & 0.9761 \\
    \hline

	\end{tabular}	
	}
	\caption{Average performance and $\Phi$ metric values over the executed (batched) problem sizes for each parallel-prefix algorithm and methodology.}
	\label{performance}	
 \vspace{-4mm}
	\end{table}

\subsection{Tridiagonal Systems Solvers Performance Results}

In the context of tridiagonal systems, the data performance is quantified in MRows/s, utilizing the formula: $N \cdot b \cdot 10^{-6} /t$, where each equation operates on four coefficients. Table \ref{performance} presents the achieved performance for the CR, PCR, LF, and WM implementations in the BPLG library using the two proposed methodologies. Furthermore, the $\Phi$ metric is computed with the optimal configuration obtained through an exhaustive search. The results demonstrate that both methodologies yield comparable performance, closely approaching the optimal configuration identified by the exhaustive search. The analytical model developed for these computation patterns exhibits excellent performance, while the ML-based search also proves highly effective, as few combinations are valid with high problem sizes requiring minimal exploration to identify the global optimal solution. It is worth noting that slight variations in execution times among GPU runs account for most of the observed differences. To mitigate variability, we conducted 100 executions for each configuration, thereby reducing the impact of such fluctuations.

\begin{figure}[t!]
\centering
\includegraphics[scale=0.38]{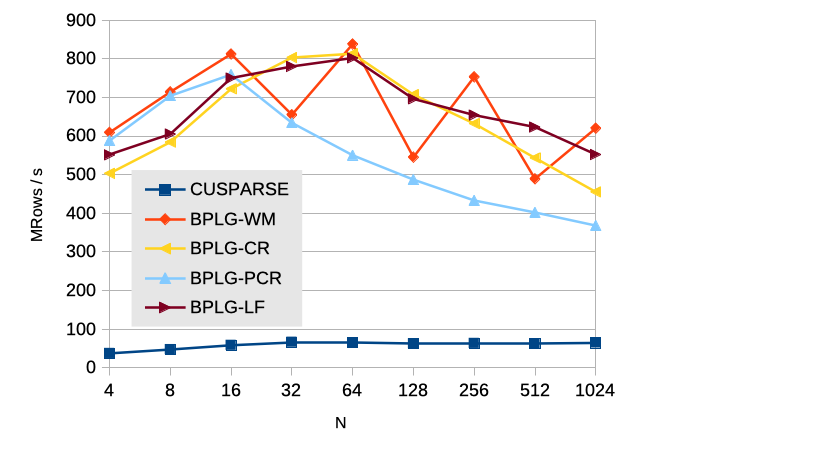}
\caption{Performance analysis for the tridiagonal solvers}
\label{ts-res}
\vspace{-4mm}
\end{figure}

Figure \ref{ts-res} illustrates the performance of the BPLG proposals in comparison to the CUSPARSE library. Notably, the jagged outline observed for \textit{BPLG-WM} stems from the CUDA implementation. When the number of elements in a thread workload (radix $r$) does not evenly divide the total number of elements ($N$), a mixed-radix technique is employed. This technique involves performing the computation in multiple steps, with a lower $r$ value for the first step and the given $r$ value for the remaining steps. Consequently, additional synchronizations are required, and the limited shared memory availability restricts the number of simultaneous threadblocks per SM. On the other hand, CUSPARSE exhibits subpar performance, which is not exclusive to the Jetson platform. Since CUDA 7.0, the performance of the CUSPARSE library for solving tridiagonal systems has experienced notable degradation.

\subsection{Scan Primitive Performance Results}

\begin{figure}[t!]
\centering
\includegraphics[scale=0.35]{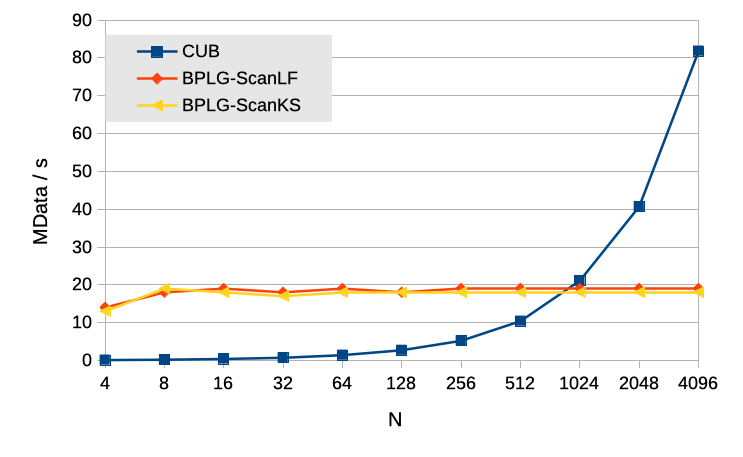}
\caption{Performance analysis for the scan operation}
\label{scan-res}
\vspace{-4mm}
\end{figure}

The performance evaluation of the Scan primitive is conducted in terms of MData/s, using the formula $N \cdot b \cdot 10^{-6}/t$. Table \ref{performance} demonstrates that both methodologies suggest configurations that align with the global optimum. Figure \ref{scan-res} provides an analysis of the performance of tuning BPLG implementations in comparison to the \textit{CUB} library. Notably, the BPLG implementations exhibit similar performance characteristics, maintaining a constant amount of shared memory regardless of the problem size $N$. This is accomplished through the utilization of \textit{shuffle} instructions for interthread communications, ensuring consistent performance across different problem sizes.

The default implementation of the \textit{CUB} library does not inherently support \textit{multi-batch} execution. However, it can be adapted by applying a segmented-scan transformation technique \cite{Sengupta06awork-efficient}. Nonetheless, invoking the library multiple times yielded the best performance in our experiments. Therefore, in the interest of fairness, the experimental results were obtained by invoking the \textit{CUB} library $b$ times, resulting in an average performance of $14.97$ MData/s. When compared to the BPLG implementations, BPLG outperforms CUB for problem sizes smaller than $N=2048$. This behavior aligns with expectations, as the number of library invocations corresponds to the number of executed batches. With a greater number of smaller problem sizes, there are more library invocations, incurring additional overhead. Conversely, for larger problem sizes, the number of batches decreases, leading to improved performance. Despite the peak of performance for large problem sizes, BPLG implementations obtain an average speed-up of $1.22x$ over the CUB library.

\subsection{Fast Fourier Transform (FFT) Performance Results}

\begin{figure}[t!]
\centering
\includegraphics[scale=0.33]{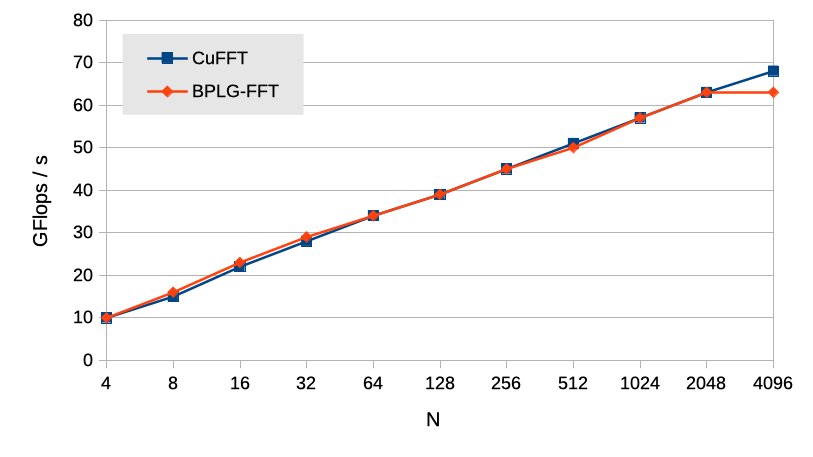}
\caption{Performance analysis for the complex FFT operation}
\label{fft-res}
\vspace{-4mm}
\end{figure}

The performance evaluation of the complex FFT is quantified in \textit{GFlops/s} using the well-established formula: $5N \cdot log_2(N) \cdot b \cdot 10^{-9} /t$, where $N$ represents the input size, $b$ denotes the number of processed batches, and $t$ is the time measured in seconds. The obtained performance results for each tuning methodology are summarized in Table \ref{performance}, showcasing competitive configurations suggested by both methodologies. Figure \ref{fft-res} provides a comprehensive comparison between the BPLG implementation and the cuFFT library in terms of performance. It is evident that the tuning BPLG version achieves highly comparable performance to cuFFT for problem sizes up to $N=2048$. However, for larger problem sizes, the shared memory limitation becomes the primary bottleneck for the BPLG version. Notably, while this Maxwell architecture offers an increased shared memory capacity of 64 KB per SM, each threadblock can only utilize up to 48 KB. Consequently, a performance drop is observed beyond $N>2048$. It is important to note that despite this decline, the cuFFT library maintains comparable performance on average with the BPLG-FFT implementation.

\subsection{Larger Problem-Size Performance Results: FFT}

%TODO: Say new batched?

\begin{figure}[t!]
\centering
\includegraphics[scale=0.33]{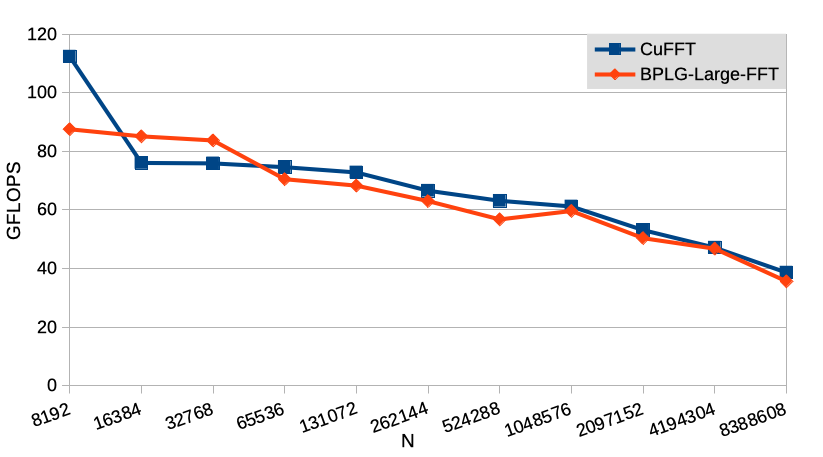}
\caption{Performance analysis for large FFTs}
\label{large-fft-res}
\vspace{-4mm}
\end{figure}

In the context of small to medium problem sizes, the analytical model has consistently demonstrated comparable or superior performance compared to the ML-based search. This finding emphasizes the pragmatic advantage of utilizing the analytical model, as it eliminates the need for costly evaluations of numerous candidates to train a surrogate model.  The ML-based search may not have appeared as a favorable option due to the relatively small search spaces, making exhaustive search or analytical modeling more practical. However, as depicted in Table \ref{performance}, a notable shift in behavior emerges for larger problem sizes. This discrepancy can be attributed to the increased complexity of the optimization problem, where up to three kernels require tuning, leading to a considerably larger search space. Consequently, the ML-based search becomes a more viable and beneficial approach.  Figure \ref{large-fft-res} shows the 
performance of the tuned BPLG implementation compared to cuFFT, where comparable performance is achieved. The highest difference in performance happens at $N=8192$, for which cuFFT launches only one kernel while BPLG launches two. This evidences the overhead associated to increase the number of kernels in a computation. The growing trend in performance along sizes observed for small and medium problem sizes is reserved for large problem sizes, as shared memory becomes a limiting factor due to its reduced capacity and more frequent high-latency global-memory accesses.

\section{Conclusions}
\label{conclusions}

%TODO: single-rank, isngle GPU, single kernel or non sparse, regular.. easy but othercases ML seems to be.
%TODO: online vs offline
%TODO: other libraries approach well but como tuning. Varianza en resultados como minimizarla

In this study, we addressed the challenge of performance portability and tuning in GPU-embedded systems, with a particular focus on the NVIDIA Jetson TX1 platform that remains a significant target for embedded systems, serving as a proof of concept for optimizations that can be extrapolated to more recent architectures. In addition to this, our investigation aimed to fill the existing gap in performance studies for parallel computation patterns in GPU-embedded systems.

To achieve performance portability, we developed and compared two tuning methodologies: an analytical model-driven approach and a Machine Learning (ML)-based approach. We examined three essential parallel computing components: the scan primitive, tridiagonal system solvers, and complex Fast Fourier Transform (FFT). These components represent performance-critical operations in numerous computing applications. Our results demonstrated that both tuning methodologies yielded highly competitive configurations, showcasing the effectiveness of each approach in specific scenarios. For regular computation patterns that require only one kernel invocation, the analytical model excelled and makes this heuristic the best choice for offline and online tuning due to the absence of candidate evaluations. On the other hand, the ML-based search, with its ability to treat the relationship between performance and tuning parameters as a black-box model, demonstrated excellent performance in all cases, but especially when the parameter space is large, for example due to the launch of several performance-interdependent kernels. The ML-based approach works better for online tuning when the performance gains through several application invocations can amortize the cost of few candidate evaluations. 

The performance evaluation of the scan primitive, tridiagonal system solvers, and complex FFT further solidified the effectiveness of our proposed methodologies. The BPLG library, tuned using both the analytical model-driven and ML-based approaches, exhibited performance that closely matched the optimal configuration obtained through exhaustive search. The BPLG implementations showcased comparable or even superior performance compared to state-of-the-art libraries such as CUSPARSE, CUB, and cuFFT for various problem sizes. Moving forward, further exploration with our methodologies on other irregular computation patterns, such as sparse or work-imbalanced patterns, will tackle the performance portability challenge for a broader catalogue of scientific workloads.

%Unless there are six authors or more give all authors' names; do not use  ``et al.''. Papers that have not been published, even if they have been submitted for publication, should be cited as ``unpublished'' \cite{b4}. Papers that have been accepted for publication should be cited as ``in press'' \cite{b5}. Capitalize only the first word in a paper title, except for proper nouns and element symbols.

% Bibliography
\bibliographystyle{IEEEtran}
\bibliography{egbibsample2}

\begin{thebibliography}{10}
\providecommand{\url}[1]{#1}
\csname url@samestyle\endcsname
\providecommand{\newblock}{\relax}
\providecommand{\bibinfo}[2]{#2}
\providecommand{\BIBentrySTDinterwordspacing}{\spaceskip=0pt\relax}
\providecommand{\BIBentryALTinterwordstretchfactor}{4}
\providecommand{\BIBentryALTinterwordspacing}{\spaceskip=\fontdimen2\font plus
\BIBentryALTinterwordstretchfactor\fontdimen3\font minus \fontdimen4\font\relax}
\providecommand{\BIBforeignlanguage}[2]{{%
\expandafter\ifx\csname l@#1\endcsname\relax
\typeout{** WARNING: IEEEtran.bst: No hyphenation pattern has been}%
\typeout{** loaded for the language `#1'. Using the pattern for}%
\typeout{** the default language instead.}%
\else
\language=\csname l@#1\endcsname
\fi
#2}}
\providecommand{\BIBdecl}{\relax}
\BIBdecl

\bibitem{7939053}
N.~Otterness, M.~Yang, S.~Rust, E.~Park, J.~H. Anderson, F.~D. Smith, A.~Berg, and S.~Wang, ``{An Evaluation of the NVIDIA TX1 for Supporting Real-Time Computer-Vision Workloads},'' in \emph{Proceedings of IEEE Real-Time and Embedded Technology and Applications Symposium (RTAS)}, 2017, pp. 353--364.

\bibitem{Ladner:1980:PPC:322217.322232}
R.~E. Ladner and M.~J. Fischer, ``Parallel prefix computation,'' \emph{J. ACM}, vol.~{27}, no.~4, pp. 831--838, 1980.

\bibitem{BPLGJacobo}
J.~Lobeiras, M.~Amor, and R.~Doallo, ``{BPLG: A Tuned Butterfly Processing Library for GPU Architectures},'' \emph{Int. J. Parallel Program.}, vol.~43, no.~6, pp. 1078--1102, 2015.

\bibitem{jos2018}
A.~P. Di\'{e}guez, M.~Amor, and R.~Doallo, ``Parallel prefix operations on gpu: Tridiagonal system solvers and scan operators,'' \emph{The Journal of Supercomputing}, vol.~75, no.~3, p. 1510–1523, mar 2019.

\bibitem{7336442}
M.~Su, J.~Tan, C.~Lin, J.~Ye, C.~Wang, and C.~Hung, ``{Constructing a Mobility and Acceleration Computing Platform with NVIDIA Jetson TK1},'' in \emph{Proceedings of IEEE 12th International Conference on Embedded Software and Systems}, 2015, pp. 1854--1858.

\bibitem{franciscomaster}
F.~J.~A. Rueda, ``{Optimizaci\'on de aplicaciones de procesado de se\~nales digitales empleando como plataforma hardware NVIDIA Jetson TK1},'' Master's thesis, Universidad Polit\'ecnica de Valencia, 2015.

\bibitem{7307646}
Y.~Ukidave, D.~Kaeli, U.~Gupta, and K.~Keville., ``{Performance of the NVIDIA Jetson TK1 in HPC},'' in \emph{Proceedings of 2015 IEEE International Conference on Cluster Computing}, 2015, pp. 533--534.

\bibitem{pepe}
H.~Halawa, H.~Abdelhafez, A.~Boktor, and M.~Ripeanu, ``Nvidia jetson platform characterization,'' in \emph{Proceedings of International European Conference on Parallel and Distributed Computing (EuroPar'17)}, vol. 10417, 2017, pp. 92--105.

\bibitem{algomas}
M.~Eisenbach, R.~Stricker, D.~Seichter, A.~Vorndran, T.~Wengefeld, and H.~Gross, ``{Speeding up Deep Neural Networks on the Jetson TX1},'' in \emph{Proceedings of Int. WS CAPRI at Join conf. on Neural Networks (IJCNN)}, 2017, pp. 11--22.

\bibitem{10.1145/3570604}
S.~K. Prashanthi, S.~A. Kesanapalli, and Y.~Simmhan, ``Characterizing the performance of accelerated jetson edge devices for training deep learning models,'' \emph{Proc. ACM Meas. Anal. Comput. Syst.}, vol.~6, no.~3, dec 2022.

\bibitem{9882480}
J.~Zhu, H.~Feng, S.~Zhong, and T.~Yuan, ``Performance analysis of real-time object detection on jetson device,'' in \emph{2022 IEEE/ACIS 22nd International Conference on Computer and Information Science (ICIS)}, 2022, pp. 156--161.

\bibitem{10.1145/3434770.3459729}
H.~A. Abdelhafez, H.~Halawa, K.~Pattabiraman, and M.~Ripeanu, ``Snowflakes at the edge: A study of variability among nvidia jetson agx xavier boards,'' in \emph{Proceedings of the 4th International Workshop on Edge Systems, Analytics and Networking}, ser. EdgeSys '21.\hskip 1em plus 0.5em minus 0.4em\relax New York, NY, USA: ACM, 2021, p. 1–6.

\bibitem{article1234}
V.~Volkov and B.~Kazian, ``{Fitting FFT onto the G80 architecture},'' \emph{Technical Report University of California, Berkeley}, 2011.

\bibitem{YURIFFT}
{Y. Dotsenko, S.S. Baghsorkhi, B. Lloyd and N.K. Govindaraju}, ``{Auto-Tuning of Fast Fourier Transform on Graphics Processors},'' in \emph{Proc. of Principles and Practice of Parallel Programming (PPoPP'11)}, 2011, pp. 257--266.

\bibitem{NUKADA}
A.~Nukada and S.~Matsuoka, ``{Auto-tuning 3-D FFT Library for CUDA GPUs},'' in \emph{Proc. of the Conf. on High Perf. Computing Networking, Storage and Analysis (SC'09)}, 2009, pp. 1--10.

\bibitem{NUKADA2}
A.~Nukada, K.~Sato, and S.~Matsuoka, ``{Scalable Multi-GPU 3-D FFT for TSUBAME 2.0 Supercomputer},'' in \emph{Proc. of the International Conference on High Performance Computing, Networking, Storage and Analysis (SC'12)}, 2012, pp. 44:1--44:10.

\bibitem{Park13}
J.~Park, G.~Bikshandi, K.~Vaidyanathan, P.~T.~P. Tang, P.~Dubey, and D.~Kim, ``{Tera-scale 1D FFT with Low-communication Algorithm and Intel\&Reg Xeon Phi\&Trade Coprocessors},'' in \emph{Proc. of the International Conference on High Performance Computing, Networking, Storage and Analysis (SC'13)}, 2013, pp. 34:1--34:12.

\bibitem{6755214}
D.~Takahashi, ``{Implementation of Parallel 1-D FFT on GPU Clusters},'' in \emph{Proc. of the IEEE 16th International Conference on Computational Science and Engineering (ICCS'13)}, 2013, pp. 174--180.

\bibitem{9555937}
B.~Li, S.~Cheng, and J.~Lin, ``tcfft: A fast half-precision fft library for nvidia tensor cores,'' in \emph{2021 IEEE International Conference on Cluster Computing (CLUSTER)}, 2021, pp. 1--11.

\bibitem{Davidson:2011:RPF}
A.~Davidson and J.~D. Owens, ``{Register Packing for Cyclic Reduction},'' in \emph{Proceed. of the 4th Workshop on General Purpose Processing on Graphics Processing Units GPGPU-4}, 2011, pp. 4:1--4:6.

\bibitem{DBLP:conf/icpp/KimWCH11}
H.~Kim, S.~Wu, L.~Chang, and W.~W. Hwu, ``{A Scalable Tridiagonal Solver for {GPUs}},'' in \emph{Proceedings of the International Conference on Parallel Processing (ICPP'11)}, 2011, pp. 444--453.

\bibitem{Chang:2012:SNS:2388996.2389033}
L.-W. Chang, J.~A. Stratton, H.-S. Kim, and W.-M.~W. Hwu, ``{A Scalable, Numerically Stable, High-performance Tridiagonal Solver Using {GPUs}},'' in \emph{Proceedings of the International Conference on High Performance Computing, Networking, Storage and Analysis (SC'12)}, 2012, pp. 27:1--27:11.

\bibitem{rtrTridiagonales}
Y.~Zhang, J.~Cohen, and J.~D. Owens, ``Fast tridiagonal solvers on the {GPU},'' in \emph{Proceed. of the 15th ACM Symposium on Principles and Practice of Parallel Programming (PPoPP'10)}, 2010, pp. 127--136.

\bibitem{TRIDIAG4}
{A. Davidson, Y. Zhang and J.D. Owens}, ``{An Auto-tuned Method for Solving Large Tridiagonal Systems on the GPU},'' in \emph{Proc. of the 25th IEEE International Parallel and Distributed Processing Symposium (IPDPS'11)}, 2011, pp. 956--965.

\bibitem{Laszlo:2016:MAB:2956571.2830568}
E.~L\'{a}szl\'{o}, M.~Giles, and J.~Appleyard, ``{Manycore Algorithms for Batch Scalar and Block Tridiagonal Solvers},'' \emph{ACM Trans. Math. Softw.}, vol.~42, no.~4, pp. 31:1--31:36, 2016.

\bibitem{Yang:2017:PSM:3092142.3092210}
W.~Yang, K.~Li, and K.~Li, ``{A Parallel Solving Method for Block-tridiagonal Equations on CPU---GPU Heterogeneous Computing Systems},'' \emph{J. Supercomput.}, vol.~73, no.~5, pp. 1760--1781, 2017.

\bibitem{9919397}
K.~Liu and W.~Xue, ``A novel compute-efficient tridiagonal solver for many-core architectures,'' \emph{IEEE Transactions on Parallel and Distributed Systems}, vol.~34, no.~1, pp. 195--206, 2023.

\bibitem{Dotsenko:2008:FSA:1375527.1375559}
Y.~Dotsenko, N.~K. Govindaraju, P.-P. Sloan, C.~Boyd, and J.~Manferdelli, ``{Fast Scan Algorithms on Graphics Processors},'' in \emph{Proceedings of the 22Nd Annual International Conference on Supercomputing}, 2008, pp. 205--213.

\bibitem{Yan:2013:SFS:2517327.2442539}
S.~Yan, G.~Long, and Y.~Zhang, ``{StreamScan: Fast Scan Algorithms for GPUs Without Global Barrier Synchronization},'' \emph{SIGPLAN Not.}, vol.~48, no.~8, pp. 229--238, Feb. 2013.

\bibitem{CUFFT}
\emph{{CUDA CUFFT Library}}, Nvidia Corporation, 2012, \url{https://developer.nvidia.com/cufft}.

\bibitem{cusparse}
NVIDIA-Corporation, ``{CUDA CUSPARSE Library},'' \url{https://developer.nvidia.com/cusparse}, 2012, last access Nov 2018.

\bibitem{CUBlib}
\emph{{CUB Library}}, \url{http://nvlabs.github.io/cub/}, NVIDIA-Corporation, 2015.

\bibitem{KERNEL13}
J.~Ansel, S.~Kamil, K.~Veeramachaneni, J.~Ragan, J.~Bosboom, U.-M. O’Reilly, and S.~Amarasinghe, ``Opentuner: An extensible framework for program autotuning,'' \emph{Parallel Architectures and Compilation Techniques, PACT}, pp. 303--315, 08 2014.

\bibitem{KERNEL}
B.~{Van Werkhoven}, ``{Kernel Tuner: A search-optimizing GPU code auto-tuner},'' \emph{Future Generation Computer Systems}, vol.~90, pp. 347--358, 2019.

\bibitem{KERNEL16}
P.~Balaprakash, S.~Wild, and P.~Hovland, ``Can search algorithms save large-scale automatic performance tuning?'' \emph{Procedia CS}, vol.~4, pp. 2136--2145, 12 2011.

\bibitem{OPENCL30}
W.~Jia, K.~A. Shaw, and M.~Martonosi, ``Starchart: Hardware and software optimization using recursive partitioning regression trees,'' in \emph{22nd International Conference on Parallel Architectures and Compilation Techniques}, 2013, pp. 257--267.

\bibitem{OPENCL31}
Y.~Zhang, M.~Sinclair, and A.~A. Chien, ``Improving performance portability in opencl programs,'' in \emph{Supercomputing}, 2013, pp. 136--150.

\bibitem{OpenCL}
T.~L. Falch and A.~C. Elster, ``Machine learning based auto-tuning for enhanced {OpenCL} performance portability,'' in \emph{2015 {IEEE} International Parallel and Distributed Processing Symposium Workshop (IPDPS)}, 2015, pp. 1231--1240.

\bibitem{cox}
J.~Bergstra, N.~Pinto, and D.~Cox, ``Machine learning for predictive auto-tuning with boosted regression trees,'' \emph{Innovative Parallel Computing, InPar 2012}, 05 2012.

\bibitem{active}
J.~Zhang, J.~Sun, W.~Zhou, and G.~Sun, ``An active learning method for empirical modeling in performance tuning,'' in \emph{IEEE Internat. Parallel and Distributed Processing Symposium (IPDPS)}, 2020, pp. 244--253.

\bibitem{BO21}
X.~Wu, M.~Kruse, P.~Balaprakash, H.~Finkel, P.~Hovland, V.~Taylor, and M.~Hall, ``Autotuning polybench benchmarks with llvm clang/polly loop optimization pragmas using bayesian optimization,'' in \emph{2020 IEEE/ACM Performance Modeling, Benchmarking and Simulation of HPC Systems (PMBS)}, 2020, pp. 61--70.

\bibitem{gptune}
Y.~Liu, W.~M. Sid-Lakhdar, O.~Marques, X.~Zhu, C.~Meng, J.~W. Demmel, and X.~S. Li, ``Gptune: Multitask learning for autotuning exascale applications,'' in \emph{Proceedings of the 26th ACM SIGPLAN Symposium on Principles and Practice of Parallel Programming}, ser. PPoPP '21, 2021, p. 234–246.

\bibitem{Dorier_2022}
M.~Dorier, R.~Egele, P.~Balaprakash, J.~Koo, S.~Madireddy, S.~Ramesh, A.~D. Malony, and R.~Ross, ``{HPC} storage service autotuning using variational- autoencoder -guided asynchronous bayesian optimization,'' in \emph{2022 {IEEE} International Conference on Cluster Computing ({CLUSTER})}.\hskip 1em plus 0.5em minus 0.4em\relax {IEEE}, sep 2022.

\bibitem{tegra}
\emph{{The Tegra X1 Whitepaper}}, Nvidia Corporation, 2015, \url{https://international.download.nvidia.com/pdf/tegra/Tegra-X1-whitepaper-v1.0.pdf}.

\bibitem{Hockney:1965:FDS:321250.321259}
R.~W. Hockney, ``{A Fast Direct Solution of {P}oisson's Equation Using {F}ourier Analysis},'' \emph{J. ACM}, vol.~{12}, no.~1, pp. 95--113, 1965.

\bibitem{hockney1988parallel}
R.~Hockney and C.~Jesshope, \emph{Parallel Computers 2: Architecture, Programming and Algorithms}.\hskip 1em plus 0.5em minus 0.4em\relax Taylor \& Francis, 1988.

\bibitem{7397622}
A.~P. Di\'eguez, M.~Amor, and R.~Doallo, ``{New Tridiagonal Systems Solvers on GPU Architectures},'' in \emph{Proc. of the 22nd International Conference on High Performance Computing (HiPC'15)}, 2015, pp. 85--94.

\bibitem{WANG}
{X. Wang and Z.G. Mou}, ``{A divide-and-conquer method of solving tridiagonal systems on hypercube massively parallel computers},'' in \emph{Proc. of the Third IEEE Symposium on Parallel and Distributed Processing (IPDPS'91)}, 1991, pp. 810--817.

\bibitem{BO}
F.-J. Willemsen, R.~van Nieuwpoort, and B.~Van~Werkhoven, ``{Bayesian Optimization for auto-tuning GPU kernels},'' in \emph{2021 International Workshop on Performance Modeling, Benchmarking and Simulation of HPC Systems (PMBS)}, 2021.

\bibitem{gptune13}
R.~Howarth, ``Mining geostatistics,'' \emph{Mineralogical Magazine}, vol.~43, pp. 563--564, 12 1979.

\bibitem{BO34}
J.~Mo{\v{c}}kus, ``On bayesian methods for seeking the extremum,'' in \emph{Optimization Techniques IFIP Technical Conference Novosibirsk}, 1975, pp. 400--404.

\bibitem{7970194}
A.~P. Di\'eguez, M.~Amor, J.~Lobeiras, and R.~Doallo, ``{Solving Large Problem Sizes of Index-Digit Algorithms on GPU: FFT and Tridiagonal System Solvers},'' \emph{IEEE Transactions on Computers}, vol.~67, no.~1, pp. 86--101, 2018.

\bibitem{pennycook}
J.~Pennycook, J.~Sewall, and V.~Lee, ``A metric for performance portability,'' \emph{eprint arXiv:1611.07409}, p.~7, 11 2016.

\bibitem{Sengupta06awork-efficient}
S.~Sengupta, A.~E. Lefohn, and J.~D. Owens, ``A work-efficient step-efficient prefix sum algorithm,'' \emph{Workshop on Edge Computing Using New Commodity Architectures}, pp. 26--27, 2006.

\end{thebibliography}

\end{document}